\documentclass[12pt]{article}
\usepackage{authblk}
\usepackage{amsthm,amsmath,amsfonts,amssymb}
\usepackage{graphicx}
\usepackage[sectionbib]{natbib}
\usepackage{url} 
\usepackage[colorlinks,citecolor=blue,urlcolor=blue]{hyperref}
\usepackage{xcolor}
\usepackage{inputenc}
\usepackage{enumerate}
\usepackage{bm}
\usepackage{appendix}
\usepackage{dsfont}
\usepackage{subfigure}
\usepackage{algorithmicx,algorithm,algpseudocode}
\usepackage{multirow}
\usepackage{mathrsfs}
\usepackage{textcomp}
\usepackage{manyfoot}
\usepackage{booktabs}
\usepackage{listings}
\usepackage{verbatim}
\usepackage{threeparttable}

\theoremstyle{plain}
\newtheorem{theorem}{Theorem}
\newtheorem{lemma}[theorem]{Lemma}
\newtheorem{remark}[theorem]{Remark}

\newcommand{\bx}{x}
\newcommand{\by}{y}
\newcommand{\ep}{\varepsilon}
\newcommand{\bep}{\varepsilon}
\newcommand{\bB}{B}
\newcommand{\bW}{W}
\newcommand{\bL}{L}
\newcommand{\bS}{S}

\newcommand{\X}{X}

\newcommand{\D}{D}
\newcommand{\I}{I}
\newcommand{\btheta}{\theta}
\newcommand{\bbeta}{\beta}

\newcommand{\hB}{\widehat{\beta}}
\newcommand{\tB}{\widetilde{\beta}}
\newcommand{\RR}{\mathbb{R}}
\newcommand{\nnz}{\mathrm{nnz}}
\newcommand{\tr}{\operatorname{tr}}
\newcommand{\var}{\operatorname{var}}
\newcommand{\pr}{\operatorname{pr}}
\newcommand{\med}{\operatorname{med}}
\newcommand{\cI}{\mathcal{I}}
\newcommand{\cO}{\mathcal{O}}
\newcommand{\cB}{\mathcal{B}}

\begin{document}

\def\spacingset#1{\renewcommand{\baselinestretch}%
{#1}\small\normalsize} \spacingset{1}


\title{\bf Core-Elements for \\
Large-Scale Least Squares Estimation}

\author[1]{Mengyu Li}
\author[2]{Jun Yu\thanks{Joint first author}}
\author[1]{Tao Li}
\author[3]{Cheng Meng\thanks{Corresponding author, chengmeng@ruc.edu.cn}}

\affil[1]{Institute of Statistics and Big Data, Renmin University of China, Beijing, China}
\affil[2]{School of Mathematics and Statistics, Beijing Institute of Technology, Beijing, China}
\affil[3]{Center for Applied Statistics, Institute of Statistics and Big Data, Renmin University of China, Beijing, China}
\date{}

\maketitle

\bigskip
\begin{abstract}
The coresets approach, also called subsampling or subset selection, aims to select a subsample as a surrogate for the observed sample and has found extensive applications in large-scale data analysis. Existing coresets methods construct the subsample using a subset of rows from the predictor matrix. Such methods can be significantly inefficient when the predictor matrix is sparse or numerically sparse. To overcome this limitation, we develop a novel element-wise subset selection approach, called core-elements, for large-scale least squares estimation. We provide a deterministic algorithm to construct the core-elements estimator, only requiring an $O(\mathrm{nnz}(X)+rp^2)$ computational cost, where $X$ is an $n\times p$ predictor matrix, $r$ is the number of elements selected from each column of $X$, and $\mathrm{nnz}(\cdot)$ denotes the number of non-zero elements. Theoretically, we show that the proposed estimator is unbiased and approximately minimizes an upper bound of the estimation variance. We also provide an approximation guarantee by deriving a coresets-like finite sample bound for the proposed estimator. To handle potential outliers in the data, we further combine core-elements with the median-of-means procedure, resulting in an efficient and robust estimator with theoretical consistency guarantees. Numerical studies on various synthetic and real-world datasets demonstrate the proposed method's superior performance compared to mainstream competitors.
\end{abstract}

\noindent%
{\it Keywords:} Coresets, Linear model, Sparse matrix, Subset selection.
\vfill

\newpage
\spacingset{1.5}

\section{Introduction}\label{sec:intro}
Sparse matrices are matrices in which most of the elements are zero.
Such matrices are common in various areas, including medical research, bioinformatics, privacy-preserving analysis, and distributed computing \citep{davis2011university, nguyen2023structure, liu2021privacy, kairouz2021advances}.
In these areas, data are usually of high sparsity due to technical noises, data privacy concerns, and transmission cost, among others \citep{konevcny2016federated, andrews2021tutorial}.
One example is the single-cell RNA-sequencing (scRNA-seq) data containing information about the gene expression level of single cells.
Owing to technical noises and intrinsic biological variability, scRNA-seq data expressed in count matrices always possess significant sparsity, known as the zero-inflation phenomenon \citep{nguyen2023structure}.
Another example is the word occurrence matrix, whose elements are calculated by multiplying two metrics, i.e., how many times a word appears in a document, and the inverse document frequency of the word across a set of documents. Such matrices are also highly sparse, especially for a short document and a large language model that contains millions of words \citep{qaiser2018text}.

In reality, many sparse matrices also exist due to missing values, i.e., the elements are not fully observed.
Numerous methods have been developed to deal with the missing values in such cases, and most of these methods aim to fill the sparse matrix with some estimated non-zero values \citep{cai2010singular, van2011mice, hastie2015matrix, muzellec2020missing}.
In this paper, however, we are less concerned with missing values and instead focus on the case that the sparse matrix itself is fully observed.

We consider large-scale data analysis where the predictor matrix is highly sparse.
One widely-used technique for large-scale data analysis is the coresets method, also called subsampling or subset selection.
These methods select a subsample as a surrogate for the observed sample.
Recently, such methods have been used pervasively in data reduction, measurement-constrained analysis, and active learning \citep{li2021modern, meng2021lowcon,settles2012active}.
Various coresets methods have been proposed for linear regression \citep{dasgupta2009sampling, boutsidis2013near,ma2015leveraging,meng2017effective,derezinski2018leveraged,ma2020asymptotic,wang2021orthogonal}, generalized linear regression \citep{wang2018optimal,ai2020optimal,ai2021optimal,yu2022optimal}, streaming time series~\citep{xie2019online,li2019online}, large-scale matrix approximation \citep{wang2013improving,alaoui2015fast,wang2019scalable}, nonparametric regression \citep{ma2015efficient,meng2020more,sun2021asymptotic,meng2022smoothing,dai2023subsampling}, among others.
Another avenue for handling large-scale data is data averaging~\citep{wang2023fast}, which is beyond the scope of this paper.

Despite the wide application, most existing coresets methods mainly focus on dense predictor matrices, and may be inefficient when the predictor matrix is of high sparsity.
In particular, most of these methods construct the subsample using certain rows from the observed sample.
When the observed predictor matrix is sparse, the selected subsample matrix also tends to be sparse for the subsampling methods that preserve the empirical distribution of the full data~\citep{mak2018support, joseph2022split, vakayil2022data}.
Such a subsample thus may lead to inefficient results, since the selected zero-valued elements have almost no impact on the down-streaming analysis, such as model estimation, prediction, and inference.
Another category of subsampling approaches is designed for prediction purposes~\citep{joseph2021supervised, chang2023predictive, dai2023subsampling}, whose resulting subsample may not be statistically similar to the full data, thereby potentially overcoming the inefficiency discussed above. Nevertheless, this class of methods still encounters the probability of selecting a sparse subsample matrix when the full predictor matrix has an extremely high sparsity.
Consequently, more efficient statistical tools suitable for sparse matrices are still meager.

In this paper, we bridge this gap by developing a novel element-wise subset selection method, called core-elements, for large-scale least squares estimation. Different from existing coresets methods that aim to select $r$ rows from the predictor matrix $\X\in\RR^{n\times p}$ $(n\gg p)$, we aim to construct a sparser subdata matrix $\X^*\in\RR^{n\times p}$ by keeping $rp$ elements of $\X$ and zeroing out the remaining elements.
Loosely speaking, our approach generalizes the existing coresets methods by getting rid of the requirement that the selected $rp$ elements have to be located in $r$ rows. Our major contributions are three-fold as follows.

\begin{enumerate}[(1)]
\item We provide a deterministic algorithm to construct the core-elements estimator for linear regression. Utilizing such an estimator, we can approximate the least squares estimation within $O(\nnz(\X)+rp^2)$ computational time, where $\nnz(\cdot)$ denotes the number of non-zero elements. Theoretical analysis demonstrates that our proposed estimator is unbiased and approximately minimizes an upper bound on estimation variance.

\item We establish a coresets-like finite sample bound for the proposed estimator with approximation guarantees.
In particular, we show that to achieve an $(1+\epsilon)$-relative error, the proposed estimator requires a subdata matrix $\X^*$ for the predictor matrix $\X$ such that the ratio $\|\X-\X^{*}\|_{2}/\|\X\|_{2}$ is $O(\epsilon^{1/2})$.
Intuitively, such a result indicates that when $\X$ gets (numerically) sparser, fewer elements are required in $\X^*$ to achieve the $(1+\epsilon)$-relative error.

\item To handle potential outliers in the data, we develop a robust variant of the core-elements method by integrating it with the widely adopted median-of-means procedure \citep{lugosi2019regularization, lecue2020robust, huang2023deepmom}.
We further propose an algorithm to construct the robust estimator within $O(\nnz(\X)+rp^2)$ time.
Theoretically, we show that the robust estimator is consistent with the true coefficient under certain regularity conditions.
\end{enumerate}

To evaluate the empirical performance and computational efficiency of the proposed strategy, we compare it with mainstream competitors through extensive synthetic and real-world datasets, including uncorrupted and corrupted data with dense or sparse predictor matrices.
Interestingly, although its primary design aims at sparse matrices, numerical findings reveal that the proposed estimator significantly outperforms its competitors regarding both estimation accuracy and CPU time, even when the predictor matrix is dense.

The remainder of this paper is organized as follows. We start in Section~\ref{sec:background} by introducing the linear model and the state-of-the-art subsampling methods. In Section~\ref{sec:model}, we develop the core-elements estimator and present its theoretical properties. The robust version of the core-elements estimator is introduced in Section~\ref{sec:mom}.
We examine the performance of the proposed estimators through extensive simulations and a real-world example in Sections~\ref{sec:simu} and~\ref{sec:real}, respectively. Technical proofs, additional experimental results, and implementation code are provided in supplementary materials.

\section{Background}\label{sec:background}
We adopt the common convention of using uppercase letters for matrices and lowercase letters for vectors or scalars. 
We denote the Euclidean norm and $\ell_1$ norm of a vector $\bx$ as $\|\bx\|$ and $\|\bx\|_1$, respectively.
For a matrix $\X$, we represent its spectral norm (i.e., largest singular value) as $\|\X\|_2$ and its Frobenius norm as $\|\X\|_F$. The condition number of $\X$, i.e., the ratio of its largest and smallest singular values, is denoted as $\kappa(\X)$. Additionally, we use the notations $E(\cdot)$, $\pr(\cdot)$, and $\tr(\cdot)$ for mathematical expectation, probability measure, and trace, respectively.

\subsection{Subsampling Methods for Least Squares Estimation}
Consider the linear regression model,
\begin{equation}\label{LM1}
y_{i} = \bx_{i}^{\top} \bbeta + \ep_{i}, \quad i=1, \ldots, n.
\end{equation}
Here $\{y_{i}\}_{i=1}^{n} \subset \RR$ are the responses, $\{\bx_{i}\}_{i=1}^{n} \subset \RR^{p}$ are the observations, $\bbeta \in \RR^{p}$ is a vector of unknown coefficients, and $\{\ep_{i}\}_{i=1}^{n}$ are independent and identically distributed (i.i.d.) error terms with zero mean and constant variance $\sigma^{2}$.
Let $\by = (y_1, \ldots, y_n) \in \RR^{n}$ be the response vector, $\X = (\bx_1, \ldots, \bx_n)^\top \in \RR^{n \times p}$ be the predictor matrix, and $\bep = (\ep_1, \ldots, \ep_n) \in \RR^{n}$ be the noise vector.
In this study, we assume that $n \gg p$, $p$ is fixed, and $\X$ is of full column rank. 
We focus on the estimation of slope parameters and assume that the full data have been centralized.
It is widely known that the ordinary least squares (OLS) estimator of $\bbeta$ takes the form
\begin{equation}\label{OLS1}
\hB_{\rm OLS} = (\X^{\top} \X)^{-1} \X^{\top} \by.
\end{equation}

In practice, the calculation of the least squares problem may suffer from high computational costs.
Specifically, standard computation of the formulation~\eqref{OLS1} requires $O(np^2)$ computational time, which can be considerable when both $n$ and $p$ are large.
To tackle the computational burden, various subsampling methods have been proposed.
The main idea of subsampling methods can be described as follows: given a predictor matrix $\X \in \RR^{n \times p}$, subsample $r$ rows (i.e., $r$ observations) from $\X$ to construct a much smaller matrix $\widehat{\X} \in \RR^{r \times p}$, and then use it as a surrogate for $\X$ in down-streaming analysis.

Most of the existing subsampling methods can be divided into two classes, i.e., the randomized subsampling approach and the design-based subsampling approach.
The former class aims to carefully design a data-dependent non-uniform sampling probability distribution such that more informative data points will be selected with larger sampling weights \citep{ma2015statistical, meng2017effective, knight2018subsampling, ma2020asymptotic, ai2021optimal}. In contrast, the latter class aims to construct the most effective subsample estimator based on certain optimality criteria developed in the design of experiments \citep{wang2019information, meng2021lowcon, wang2021orthogonal, wang2022sampling, chasiotis2024subdata}. Despite the numerous subsampling algorithms proposed, most of them rely on row-wise sampling, which can be less effective when dealing with (numerically) sparse predictor matrices, as discussed in Section~\ref{sec:intro}. In contrast, element-wise sampling is more adept at exploiting the inherent sparsity of data, motivating the development of the core-elements method. See Table~\ref{tab:method-comparison} for a comprehensive comparison. We also refer readers to \cite{li2021modern} and \cite{yu2023review} for recent reviews.

\begin{table}[!t]
\centering
\caption{Comparison of mainstream subsampling methods on computational cost, method type, and sampling type.}
\label{tab:method-comparison}
\begin{tabular}{lccc}
\toprule
Method & Computational cost\textsuperscript{$\ast$} & Method type & Sampling type \\
\midrule
Full sample & $O(np^2)$ & - & - \\
Uniform subsampling & $O(rp^2)$ & Randomized & Row-wise \\
Doubly sketching & \multirow{2}{*}{$O(rp^2)$} & \multirow{2}{*}{Randomized} & \multirow{2}{*}{Row-wise} \\
\citep{hou2023generalized} & & & \\
Leverage score subsampling & \multirow{2}{*}{$O(np^2 + rp^2)$\textsuperscript{$\dagger$}} & \multirow{2}{*}{Randomized} & \multirow{2}{*}{Row-wise} \\
\citep{ma2015leveraging} & & & \\
IBOSS & \multirow{2}{*}{$O(np + rp^2)$} & \multirow{2}{*}{Deterministic} & \multirow{2}{*}{Row-wise} \\
\citep{wang2019information} & & & \\
OSS & \multirow{2}{*}{$O(np \log r + rp^2)$} & \multirow{2}{*}{Deterministic} & \multirow{2}{*}{Row-wise} \\
\citep{wang2021orthogonal}  & & & \\
D-optimal subsampling & \multirow{2}{*}{$O(np^2 + rp^2)$\textsuperscript{$\ddagger$}} & \multirow{2}{*}{Deterministic} & \multirow{2}{*}{Row-wise} \\
\citep{reuter2023d} & & & \\
Core-elements (proposed) & $O(\text{nnz}(X) + rp^2)$ & Deterministic & Element-wise \\
\bottomrule
\end{tabular}
\begin{tablenotes}
\footnotesize
\item \textsuperscript{$\ast$} Here, each method subsamples $r$ rows or $s=rp$ elements from an $n \times p$ predictor matrix.
\item \textsuperscript{$\dagger$} The $O(np^2)$ component can be reduced to $O(np\log n)$ by involving some random projection-based approximation methods \citep{drineas2012fast}.
\item \textsuperscript{$\ddagger$} The $O(np^2)$ component can be reduced to $O(np)$ by using a simplified approximation that ignores non-diagonal elements of the covariance matrix.
\end{tablenotes}
\end{table}

Closely related to the subsampling methods are the coresets methods.
These methods aim to select a subsample in a deterministic way, such that a loss function $\mathcal{L}$ based on the subsample estimator is bounded by the loss function based on the full-sample estimator multiplying a constant $(1+\epsilon)$ \citep{boutsidis2013near, munteanu2018coresets, feldman2020turning, maalouf2022unified}.
In linear models, a subsample $\widehat{\X}$ is called an $(1+\epsilon)$-coreset ($\epsilon>0$), if there exists an estimator $\tB$ constructed by $\widehat{\X}$, such that 
\begin{equation*}
\|\by-\X\widehat{\bbeta}_{\rm OLS}\|^2\leq  \|\by-\X\tB\|^2\leq (1+\epsilon)\|\by-\X\widehat{\bbeta}_{\rm OLS}\|^2.
\end{equation*}

\subsection{Sparse and Numerically Sparse Matrices}
Recall that sparse matrices are matrices in which most elements are zero.
Commonly, the sparsity is measured by using the $\ell_0$ norm (i.e., the number of non-zero elements).
Such a procedure, however, may not accurately reflect the simplicity of nearly sparse instances with a large number of small but non-zero elements. 
To combat the obstacle, existing literature also considers the so-called numerically sparse matrices \citep{gupta2018exploiting, braverman21near}.
Intuitively, numerically sparse is a weaker condition than sparse in which we do not require most elements to be zero, only that most elements are small enough to be ignored.
Examples of numerically sparse are widely encountered in practice, including but not limited to linear programming constraints of the form $x_1 \ge \sum_{i=1}^n x_i/n$, and physical models whose strength of interaction decays with distance \citep{carmon2020coordinate}.

\subsection{Element-wise Sampling on Sparse Data Matrices}
Consider the scenario that the observed predictor matrix $\X\in\RR^{n\times p}$ is a (numerically) sparse matrix in which most of the elements are (nearly) zero.
Oftentimes, of interest is to find a sparser matrix $\X^+\in\RR^{n\times p}$ that is a good proxy for $\X$.
Here, $\X^+$ is also a sparse matrix, from which the non-zero elements are a subset of the non-zero elements with respect to (w.r.t.) $\X$.
The problem of finding such a matrix $\X^+$ has many applications in eigenvector approximation \citep{arora2006fast, achlioptas2007fast, el2010second, kundu2017recovering, gupta2018exploiting}, semi-definite programming \citep{arora2005fast, d2011subsampling, garber2016sublinear}, matrix completion \citep{candes2009exact, candes2010power, chen2014coherent}, optimal transport problems \citep{li2023importance, li2023efficient, hu2024sampling}, and nonparametric regression \citep{li2024nonparametric}, among others.

To construct such a matrix $\X^+$, most existing methods aim to design a good sampling probability distribution $p_{ij}$ ($i=1,\ldots,n$; $j=1,\ldots,p$), such that the approximation error $\|\X-\X^+\|_{2}$ is as small as possible given a sampling budget $s$. 
\cite{achlioptas2007fast} proposed the so-called $\ell_{2}$ sampling such that $p_{ij} \propto x_{ij}^{2}$, which was later refined by \cite{drineas2011note}.
Alternatively, \cite{arora2006fast} proposed $\ell_{1}$ sampling, where $p_{ij} \propto |x_{ij}|$. 
Later, \cite{achlioptas2013near} proposed a near-optimal probability distribution using matrix-Bernstein inequality. 
Their approach can be regarded as a combination of two $\ell_{1}$-based distributions: $p_{i j} \propto |x_{ij}|$ when $s$ is small, and $p_{i j} \propto |x_{ij}|\cdot \|\bx_i\|_{1}$ as $s$ grows.
More recently, \cite{kundu2017recovering} developed hybrid-$(\ell_{1}, \ell_{2})$ sampling, which is also a convex combination of probabilities previously proposed.

In general, existing literature on element-wise sampling mainly focus on the algorithmic perspective, such that the applications of interest are usually compressed sensing and matrix recovery. 
Nevertheless, element-wise sampling with a statistical perspective, such that the applications involve providing effective and efficient solutions for large-scale statistical modeling and inference, is still a blank field and remains further studied.

\section{Core-Elements}\label{sec:model}
In this section, we present our main algorithm and the theoretical properties of the proposed estimator. We first introduce the definition of core-elements and then construct an unbiased core-elements estimator. Following this, we derive an upper bound for the variance of such an estimator by utilizing the matrix-form Taylor expansion. Subsequently, we provide an algorithm for core-elements selection, resulting in an estimator that approximately minimizes this upper bound. Furthermore, a coresets-like finite sample bound is provided to quantify the approximation error for the proposed estimator.

\subsection{Problem Setup}
We consider the problem of large-scale least squares estimation when the predictor matrix $\X$ is sparse or numerically sparse.
To utilize the sparse structure, we propose to construct a sparser subdata matrix $\X^*\in\RR^{n\times p}$ from $\X$ carefully, and use $\X^*$ to construct an efficient least squares approximation. 
In particular, given a positive integer $s (< np)$, let $\bS$ be an $n\times p$ matrix such that its elements involve $s$ ones and $np-s$ zeros.
The subdata matrix $\X^*$ then can be formulated as $\X^* = \bS\odot \X$, where $\odot$ represents the Hadamard product, i.e., element-wise product.
In other words, $\X^*$ is produced by keeping $s$ elements of $\X$ and zeroing out the remaining elements.
We then propose a general estimator that takes the form
\begin{equation}\label{eqn1}
    \tB(\D) = \D\X^{\ast \top}\by, 
\end{equation}
where $\D\in\RR^{p\times p}$ is a scaling matrix to be determined.
For simplicity, suppose that both $\X^{\top} \X$ and $\X^{\ast \top} \X$ are invertible in context; otherwise, their inverse operation should be replaced by a generalized inverse.

A natural question arises: given a fixed budget $s$, among all the estimators that take the form~\eqref{eqn1}, how to construct the scaling matrix $\D$ and the sparse subdata matrix $\X^*$, such that (i) the estimator $\tB(\D)$ is unbiased and (ii) its estimation variance is as small as possible?
For the first part of this question, one can see that when $\D = (\X^{\ast \top}\X)^{-1}$, $E \{\tB(\D)\} = (\X^{\ast \top} \X)^{-1} \X^{\ast \top}\{\X \bbeta + E(\bep)\}=\bbeta$,
indicating that $\tB(\D)$ is unbiased to $\bbeta$.
Such a finding motivates us to focus on the unbiased estimator 
\begin{equation*}
    \tB = (\X^{\ast \top}\X)^{-1}\X^{\ast \top}\by.
\end{equation*}
Consider the classical subsample-based estimator 
$$\tB^\prime = (\widehat{\X}^{\top}\widehat{\X})^{-1}\widehat{\X}^{\top}\widehat\by,$$
where $\widehat{\X}\in\RR^{r\times p}$ represents a subsample (of full column rank) consisting of $r$ rows from $\X$, and $\widehat\by\in\RR^{r}$ contains the corresponding elements from $\by$.
When the selected $s=rp$ elements are located in $r$ rows, i.e., when the element-wise subset selection degenerates to the row selection, it can be shown that $\tB = \tB^\prime$.
While in general cases, these two estimators are different.

Next, we consider the variance of the estimator $\tB$.
Recall that $\sigma^2$ represents the variance of random errors in model~\eqref{LM1}. 
Our goal is to find a subdata matrix $\X^{\ast}$ that minimizes the estimation variance, which has a closed form
\begin{align}
    E(\|\tB-\bbeta\|^{2}|\X)
    =& E\{\by^\top \X^{\ast} (\X^{\top} \X^\ast)^{-1} (\X^{\ast \top} \X)^{-1} \X^{\ast \top} \by|\X\} -\bbeta^\top \bbeta \nonumber\\
    =& \bbeta^\top \bbeta + \sigma^2 \tr\{\X^\ast(\X^{\top} \X^\ast)^{-1}(\X^{\ast \top} \X)^{-1} \X^{\ast \top}\} - \bbeta^\top \bbeta \nonumber\\
    =& \sigma^2\|(\X^{\ast \top} \X)^{-1} \X^{\ast \top}\|_F^2. \label{eq:tB-variance}
\end{align}
However, this variance term is challenging to minimize directly.
To overcome the obstacle, we provide an upper bound for \eqref{eq:tB-variance} and aim to minimize the upper bound instead.
The upper bound is derived by utilizing the matrix-form Taylor expansion (see Chapter 1, \cite{higham2008functions}), detailed in Lemma~\ref{lem3}.

\begin{lemma}\label{lem3}
Let $\bL = \X-\X^*$. A Taylor expansion of $E(\|\tB-\bbeta\|^{2} | \X)$ around the point $\X^*=\X$ yields the following upper bound,
\begin{align}\label{eqn3} 
E(\|\tB-\bbeta\|^{2} | \X) \leq \sigma^2 [\tr\{(\X^{\top}\X)^{-1}\}
+ \|(\X^{\top}\X)^{-1}\|_2^2 \|\bL\|_F^2]\{1+O(\lambda_0)\}. 
\end{align}
Here, assume the spectral radius $\lambda_0 = \|(\X^\top\X)^{-1}\bL^\top\X\|_2<1$ to ensure the convergence of the matrix series.
\end{lemma}

When the Taylor expansion in Lemma~\ref{lem3} is valid, the inequality~\eqref{eqn3} indicates that the upper bound of the estimation variance decreases as $\|\bL\|_F$ and the remainder $\lambda_0$ decreases.  
Considering $\lambda_0$, we have
\begin{align*}
\lambda_0 \leq \|(\X^\top\X)^{-1}\|_2 \|\X\|_2 \|\bL\|_2 \leq \|(\X^\top\X)^{-1}\|_2 \|\X\|_2 \left(p\underset{j\in\{1,\ldots,p\}}{\mbox{max}}{l}^{(j)\top}{l}^{(j)}\right)^{1/2}
,
\end{align*}
where ${l}^{(j)}$ denotes the $j$th column of $\bL$.
Such an inequality indicates that a smaller value of the maximum column norm of $\bL$ is associated with a smaller $\lambda_0$.
As a result, to minimize the upper bound in Lemma~\ref{lem3}, we need to keep both $\|\bL\|_F$ and the column norms of $\bL$ as small as possible.

Motivated by this, we propose to construct $\X^\ast$ by keeping the elements with the top largest absolute values w.r.t. each column of $\X$ and zeroing out the remaining.
Intuitively, for a fixed number of selected elements, such $\bL$ has the approximately minimum column norm respecting every column.
Thus, both the values of $\|\bL\|_F$ and $\|\bL\|_2$ will be approximately minimized, resulting in a relatively small upper bound of the estimation variance in Lemma~\ref{lem3}.
Moreover, the column-wise process can prevent any entire column of $\X$ from being discarded, thus avoiding producing a singular $\X^*$ and ensuring the estimability of coefficients.
We call such a procedure ``core-elements'' since the idea behind it is analogous to ``coresets'', except that what we select are elements instead of rows.

\subsection{Main Algorithm}
\begin{figure}[htbp]
    \centering
    \includegraphics[width=2.5in]{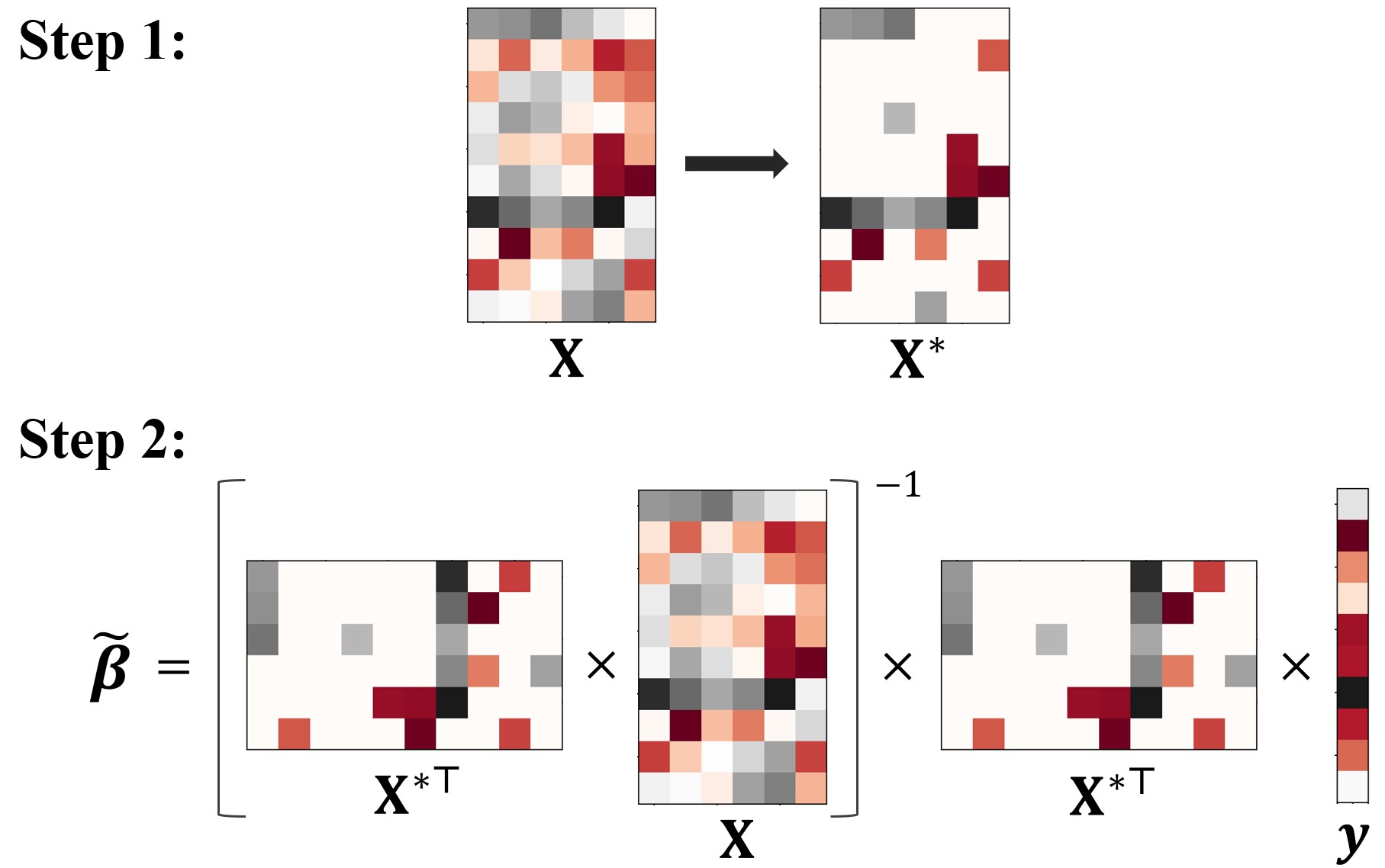}
    \caption{Illustration for Algorithm~\ref{alg:ALG1}. Each element of a matrix is labeled with a different color, such that larger positive and smaller negative values are labeled with more red and more black colors, respectively. 
    Step 1 illustrates the core-elements selection, where $r=3$ elements with the largest absolute values w.r.t. each column are selected. Step 2 illustrates the proposed estimator.}
    \label{fig:core}
\end{figure}

Without loss of generality, we assume the number of selected elements $s=rp$, where $r$ is an integer. 
Algorithm~\ref{alg:ALG1} summarizes the construction of core-elements and the proposed estimator, which are illustrated in Fig.~\ref{fig:core}.

\begin{algorithm}[ht]
\caption{\textsc{Core-elements}($\X, \by, r$)}
\begin{algorithmic}[1]
\State {\bf Input:}
$\X=(x_{ij})\in\RR^{n\times p}$,  $\by\in\RR^n$, $r\in\mathbb{Z}_{+}$ 
\State Initialize $\bS=(0) \in\RR^{n\times p}$
\For{$j=1,\ldots,p$}
\State Let $\mathcal{I} = \{i_1,\ldots,i_r\}$ be an index set, such that $\{|x_{i_q j}|\}_{q=1}^r$ are the $r$ largest ones among $\{|x_{ij}|\}_{i=1}^n$
\State Let $s_{i_q j} = 1$, $q=1,\ldots,r$
\EndFor
\State $\X^* = \bS\odot \X$, where $\odot$ represents the element-wise product
\State \textbf{Return} $\tB = (\X^{\ast \top} \X)^{-1} \X^{\ast \top} \by$
\end{algorithmic}
\label{alg:ALG1}
\end{algorithm}

Consider the computational cost of Algorithm~\ref{alg:ALG1}.
Constructing the matrix $\X^*$ using a partition-based selection algorithm requires $O(\nnz(\X))$ time \citep{musser1997introspective, martinez2004partial, wang2019information}.
Each column of $\X^*$ contains at most $r$ non-zero elements, thus calculating $\X^{*\top}\X$ takes $O(rp^2)$ time by using sparse matrix representations and operations.
The computing time for $\widetilde{\bbeta}$ is thus at the order of $O(rp^2+p^3)$. 
Therefore, the overall computational cost of Algorithm~\ref{alg:ALG1} is $O(\nnz(\X)+rp^2)$, which becomes $O(\nnz(\X))$ when $n\gg r$.
In the large-sample scenario such that $n\gg p$, such an algorithm is much faster than the popular leverage-based subsampling methods \citep{ma2015statistical}.
This is because the leverage-based methods involve the singular value decomposition of the full predictor matrix, requiring a cost of the order $O(np^2)$.

\subsection{Theoretical Properties}
We now present our main theorem, indicating that the selected elements are $(1+\epsilon)$-core-elements for least squares estimation. 
In other words, under some regularity conditions, the proposed estimate achieves the $(1+\epsilon)$-relative error w.r.t. the $\ell_2$ loss.
Technical proofs are provided in the supplementary material.

\begin{theorem}\label{th1}
Given $\X\in \RR^{n\times p}$, $\by \in \RR^{n}$ and $\epsilon > 0$, let $\X^{\ast}$ be the subdata matrix and $\widetilde{\bbeta}$ be the estimate that calculated by Algorithm~\ref{alg:ALG1}. 
When $\X^{*}$ satisfies $\|\X-\X^{*}\|_{2}\leq\epsilon^\prime\|\X\|_{2}$ with
\begin{equation}\label{eq:th1-2}
0 < \epsilon^\prime \le \frac{1}{\kappa^2(\X)} \left[1+\frac{\{\kappa^2(\X)+1\}\|\by\|}{\epsilon^{1/2} \|\by - \X \hB_{\rm OLS}\|}\right]^{-1},
\end{equation}
we have
\begin{equation}\label{eq:th1}
    \|\by-\X \hB_{\rm OLS}\|^{2} \leq \|\by-\X \widetilde{\bbeta}\|^{2} \leq (1+\epsilon)\|\by-\X \hB_{\rm OLS}\|^{2}.
\end{equation}
\end{theorem}

Theorem~\ref{th1} indicates that to achieve the $(1+\epsilon)$-relative error in inequality~\eqref{eq:th1}, Algorithm~\ref{alg:ALG1} requires a subdata matrix $\X^*$ such that the ratio $\|\X-\X^{*}\|_{2}/\|\X\|_{2}$ is $O(\epsilon^{1/2})$.
This rate is supported by empirical results in Section~\ref{sec:simu}.
Intuitively, such a result also indicates that when the predictor matrix $\X$ gets (numerically) sparser, fewer elements are required in $\X^*$ to achieve the same relative error w.r.t. the $\ell_2$ loss.

In addition, the value of $\epsilon^\prime$ also depends on the condition number $\kappa(\X)$ and the relative sum of squares error (SSE) $\|\by - \X \hB_{\rm OLS}\|^2/\|\by\|^2$. 
Specifically, a larger $\epsilon^\prime$ is admitted to achieve the $(1+\epsilon)$-relative error if the condition number $\kappa(\X)$ decreases and the SSE increases. The following remark discusses the relationship between $r$ and $\epsilon^\prime$.

\begin{remark}\label{remark1}
Suppose $\X$ is a sparse covariate matrix with $\alpha \times 100\%$ non-zero elements $(0<\alpha\leq 1)$, and each column has the same number of non-zero elements. Further, suppose the non-zero elements of $\X$ are i.i.d. from a continuous cumulative distribution function $F$. We consider two specific cases of the distribution $F$ as follows.
\begin{itemize}
    \item Case 1: uniform distribution over $(-1,1)$.
    
    If the subsample parameter $r$ in Algorithm~\ref{alg:ALG1} satisfies $r < \alpha n$ and
    \begin{equation*}\label{eq:r-uniform}
    \frac{r}{n} \geq \alpha - \frac{(\alpha \epsilon^{\prime} \|\X\|_{2})^{2/3}}{(2np)^{1/3}},
    \end{equation*}
    then the subdata matrix $\X^*$ achieves the condition $\|\X-\X^{*}\|_{2}\leq\epsilon^\prime\|\X\|_{2}$ in Theorem~\ref{th1} with high probability for a relatively large $n$. Under a general condition $\|\X\|_2=O((np)^{1/2})$, which can be satisfied as long as all the elements in $\X$ are bounded, such a result indicates that Algorithm~\ref{alg:ALG1} needs to select around $\{\alpha - (c \alpha \epsilon^{\prime})^{ 2/3}\}\times 100\%$ elements to achieve the $(1+\epsilon)$-relative error for a constant $c>0$.

    \item Case 2: standard normal distribution on $\mathbb{R}$.
    
    If $r < \alpha n$ satisfies
    \begin{equation}\label{eq:r-normal}
    \frac{r}{n} \geq \alpha - \min\left\{ \alpha\phi,  \frac{(\epsilon^{\prime} \|\X\|_{2})^2}{2 G^{-1}(\phi) np}\right\}, 
    \end{equation}
    where $0<\phi<1$ and $G$ is the cumulative distribution function of the chi-squared distribution with $1$ degree of freedom, then the condition $\|\X-\X^{*}\|_{2}\leq\epsilon^\prime\|\X\|_{2}$ holds with high probability when $n$ is sufficiently large. 
    To achieve the $(1+\epsilon)$-relative error, \eqref{eq:r-normal} indicates us to select $[\alpha - \min\{\alpha\phi,  (c\epsilon^{\prime})^{2}/G^{-1}(\phi)\}]\times 100\%$ elements under the condition $\|\X\|_2=O((np)^{1/2})$, where $c>0$ is a constant.
\end{itemize}

\end{remark}


\section{MOM Core-Elements}\label{sec:mom}
To guarantee the robustness of the core-elements approach, we modify Algorithm~\ref{alg:ALG1} by combining it with the popular median-of-means (MOM) procedure \citep{hsu2014heavy, lugosi2019regularization, lecue2019learning, lecue2020robust,  mathieu2021m, huang2023deepmom} to provide a robust version of core-elements, called ``MOM core-elements''. We first propose an algorithm in the divide-and-conquer framework to obtain the MOM core-elements estimator, and then we establish consistency of the proposed estimator under regularity conditions.

\subsection{Proposed Algorithm}
Due to the existence of outliers, we relax the standard i.i.d. setup to the $\cI \cup \cO$ framework following the work of \cite{lecue2020robust}. Specifically, we assume that data are partitioned into two (unknown) sets, $\cI$ and $\cO$, such that $\cI \cup \cO = \{1,\ldots,n\}$ and $\cI \cap \cO = \varnothing$. Data $\{(\bx_{i}, y_{i})\}_{i \in \cI}$ are i.i.d. informative data, and data $\{(\bx_{i}, y_{i})\}_{i \in \cO}$ are outliers on which no assumption is granted.

Given a random partition of the full data $(\X,\by) = \{(\bx_{i}, y_{i})\}_{i=1}^n$ into blocks of equal sizes, the principle of MOM estimator is that we first obtain an estimation on each block independently, and then we aggregate the results from these blocks by taking the median.
Let $k$ be the number of blocks and $n_l$ ($l=1,\ldots,k$) be the size of each block.
Without loss of generality, we assume that both $n$ and $r$ are divisible by $k$.
Then $n_l \equiv n/k$ for $l=1,\ldots,k$.
Denote $(\X^{(l)},\by^{(l)}) \in \RR^{n_l\times p} \times \RR^{n_l}$ be the data partitioned into the $l$th block.
For the $l$th block, we construct the core-elements matrix $\X^{(l)\ast}$ containing $r_l p$ non-zero elements, where $r_l \equiv r/k$ for $l=1,\ldots,k$. Next, we obtain the corresponding estimator $\tB^{(l)} = (\X^{(l)\ast\top} \X^{(l)})^{-1} \X^{(l)\ast\top} \by^{(l)}$. Then, the MOM core-elements estimator is defined as
\begin{equation}\label{eq:mom-estimation}
    \tB_{\rm MOM} = \med(\tB^{(1)}, \ldots, \tB^{(k)}),
\end{equation}
where $\med(\cdot)$ denotes the coordinate-wise median. 
Algorithm~\ref{alg:ALG2} details the MOM core-elements procedure. 

\begin{algorithm}[ht]
\caption{\textsc{MOM Core-elements}($\X, \by, r, k$)}
\begin{algorithmic}[1]
\State {\bf Input:}
$\X=(x_{ij})\in\RR^{n\times p}$,  $\by\in\RR^n$, $r, k \in\mathbb{Z}_{+}$ 
\State Partition $(\X, \by)$ into $k$ blocks $\{(\X^{(l)}, \by^{(l)})\}_{l=1}^k$ randomly and evenly
\For{$l=1,\ldots,k$}
\State Compute $\tB^{(l)} = \textsc{Core-elements}(\X^{(l)}, \by^{(l)}, r/k)$
\EndFor
\State \textbf{Return} $\tB_{\rm MOM} = \med(\tB^{(1)}, \ldots, \tB^{(k)})$
\end{algorithmic}
\label{alg:ALG2}
\end{algorithm}

In Algorithm~\ref{alg:ALG2}, the number of blocks $k$ indicates the robustness of our proposed estimator. A popular measure to quantify robustness is the breakdown point \citep{hampel1968contributions, donoho1983notion, donoho1992breakdown, lecue2020robust}, defined as the smallest proportion of corrupted observations needed to push an estimator to infinity.
The breakdown point of Algorithm~\ref{alg:ALG2} is $\lfloor k / 2\rfloor/n$, because less than $\lfloor k / 2\rfloor$ outliers may corrupt at most $\lfloor k / 2\rfloor$ blocks, leaving the median in~\eqref{eq:mom-estimation} equal to the estimation on a single block with uncorrupted data. 
Remark that Algorithm~\ref{alg:ALG1} is a special case of Algorithm~\ref{alg:ALG2} when $k = 1$, which is applicable to datasets without extreme outliers.

Consider the computation time of Algorithm~\ref{alg:ALG2}.
Constructing the core-elements estimator on the $l$th block requires $O(\nnz(\X^{(l)})+rp^2/k+p^3)$ time, and the aggregation step needs $O(kp)$ time. Thus, the total computational cost of Algorithm~\ref{alg:ALG2} is at the order of $O(\nnz(\X)+rp^2+kp^3+kp) = O(\nnz(\X)+rp^2)$, where the equation holds because $r/k$ should be at least of the same order as $p$ for ensuring the well-definedness of $\tB^{(1)}, \ldots, \tB^{(k)}$. 
Such cost is the same as that of Algorithm~\ref{alg:ALG1}.

\subsection{Theoretical Properties}
Now we provide the convergence of the MOM core-elements estimator. To begin with, we introduce the following regularity conditions.

\begin{enumerate}[(H1)]
    \item Denote the Fisher information matrix $\I^{(l)}=n_l^{-1}\X^{(l)\top}\X^{(l)}$ on the $l$th block for $l=1,\ldots,k$. Assume that
    $c<\inf_l\lambda_{\min}(\I^{(l)}) \le \sup_l \lambda_{\max}(\I^{(l)})<\infty$ for some constant $c>0$, where $\lambda_{\max}(\cdot)$ and $\lambda_{\min}(\cdot)$ respectively stand for the maximum and minimum eigenvalues.
    \item Let $\bL^{(l)} = \X^{(l)}-\X^{(l)\ast}$ for $l=1,\ldots,k$. Assume that $\sup_l\|\bL^{(l)}\|_F^2/n_l^2\to 0$ as $n_l\to\infty$.
    \item Assume that the spectral radius $\lambda_0^{(l)} = \|(\X^{(l)\top}\X^{(l)})^{-1}\bL^{(l)\top}\X^{(l)}\|_2 < 1$ for $l=1,\ldots,k$.
    \item Suppose that $k>2|\cB_\cO|+1$, where $\cB_\cO$ is the set of blocks containing at least one outlier and $|\cdot|$ denotes the cardinal number. Further, assume that $\cB_\cO$ contains finite elements.
\end{enumerate}

\begin{theorem}\label{th2:mom}
Suppose the conditions (H1)--(H4) hold almost surely. As $k, n_l \to \infty~(l=1,\ldots,k)$, $\tB_{\rm MOM}$ converges to $\bbeta$ in probability, i.e., for any given $\epsilon>0$, 
it holds that
\begin{equation*}
    \pr(\|\tB_{\rm MOM} - \bbeta\| > \epsilon) \to 0.
\end{equation*}
\end{theorem}

In Theorem~\ref{th2:mom}, condition (H1) is commonly assumed in the literature. Conditions (H2) and (H3) bound the error of the local core-elements estimator on each block by using Lemma~\ref{lem3}. Condition (H4) guarantees the effectiveness of the MOM procedure according to its breakdown point discussed above.

\section{Simulation Studies}\label{sec:simu}
In this section, we first evaluate the performance of core-elements (i.e., Algorithm~\ref{alg:ALG1}) in estimating $\bbeta$ and predicting $\by$ on uncorrupted synthetic datasets. Subsequently, we provide empirical evidence to support the error bound in Theorem~\ref{th1}. Next, we consider corrupted datasets to show the effectiveness and robustness of MOM core-elements (i.e., Algorithm~\ref{alg:ALG2}). Finally, we demonstrate the advantage of the proposed strategy over other subsampling methods w.r.t. computational efficiency.

\subsection{Performance on Uncorrupted Data}\label{sec:simu-uncorrupted}
We use CORE to refer to the estimator in Algorithm~\ref{alg:ALG1}.
For comparison, we consider the full sample OLS estimation (FULL) and several state-of-the-art subsampling methods mentioned in Table~\ref{tab:method-comparison}, including uniform subsampling (UNIF), doubly sketching (DOUBLY) \citep{hou2023generalized}, basic leverage subsampling (BLEV) \citep{drineas2006fast, ma2015statistical}, shrinkage leverage subsampling (SLEV) with shrinkage parameter being $0.9$ \citep{ma2015statistical}, information-based optimal subset selection (IBOSS) \citep{wang2019information}, orthogonal subsampling (OSS) \citep{wang2021orthogonal}, and D-optimal subsampling (DOPT) \citep{reuter2023d}.

For uncorrupted data, the predictor matrix $\X$ is generated from different kinds of widely-used distributions:
\begin{enumerate}[(D1)]
\item multivariate normal distribution, $N(0_p, {\Sigma})$;
\item multivariate log-normal distribution, $LN(0_p, {\Sigma})$;
\item multivariate t-distribution with 3 degrees of freedom, $t_{3}(0_p, {\Sigma})$,
\end{enumerate}
where ${\Sigma}=(\sigma_{i j}) \in \RR^{p\times p}$ is a covariance matrix with $\sigma_{i j}=0.6^{|i-j|}$ for $i,j=1,\ldots,p$.

To introduce sparsity, after generating the predictor matrix and centering it, we randomly zero out their elements with a sparsity ratio $\alpha$.
Specifically, we randomly select $\alpha\times100\%$ of the elements and set these elements to be zero.
Consider $\alpha=\{0,0.2,0.4,0.6,0.8\}$, referred to as (R1)--(R5), respectively.
Next, we add a small random perturbation following $U(-0.1,0.1)$ to each zero element of the predictor matrix to obtain a numerically sparse matrix.
Then, (R1) corresponds to a completely dense matrix, and (R5) corresponds to a highly numerically sparse matrix.
We then generate the response $\by$ from the linear model~\eqref{LM1}.
The true coefficient $\bbeta$ is a $p$-dimensional vector of ones, and the signal-to-noise ratio, defined as $\operatorname{SNR} = \var(\X\bbeta) / \sigma^2$, is set to be $4$.
The simulations in misspecified linear models and alternative choices of true $\bbeta$ are relegated to the supplementary material.
Let the sample size $n = 10^4$ and dimension $p = 10^2$.
For the row-wise subsampling methods, we select $r\in \{2p,4p,6p,8p,10p\}$ rows for each of these methods.
For a fair comparison, we select $s = rp$ elements for the proposed core-elements method.

\begin{figure}[!t]
    \centering
    \includegraphics[width=5in]{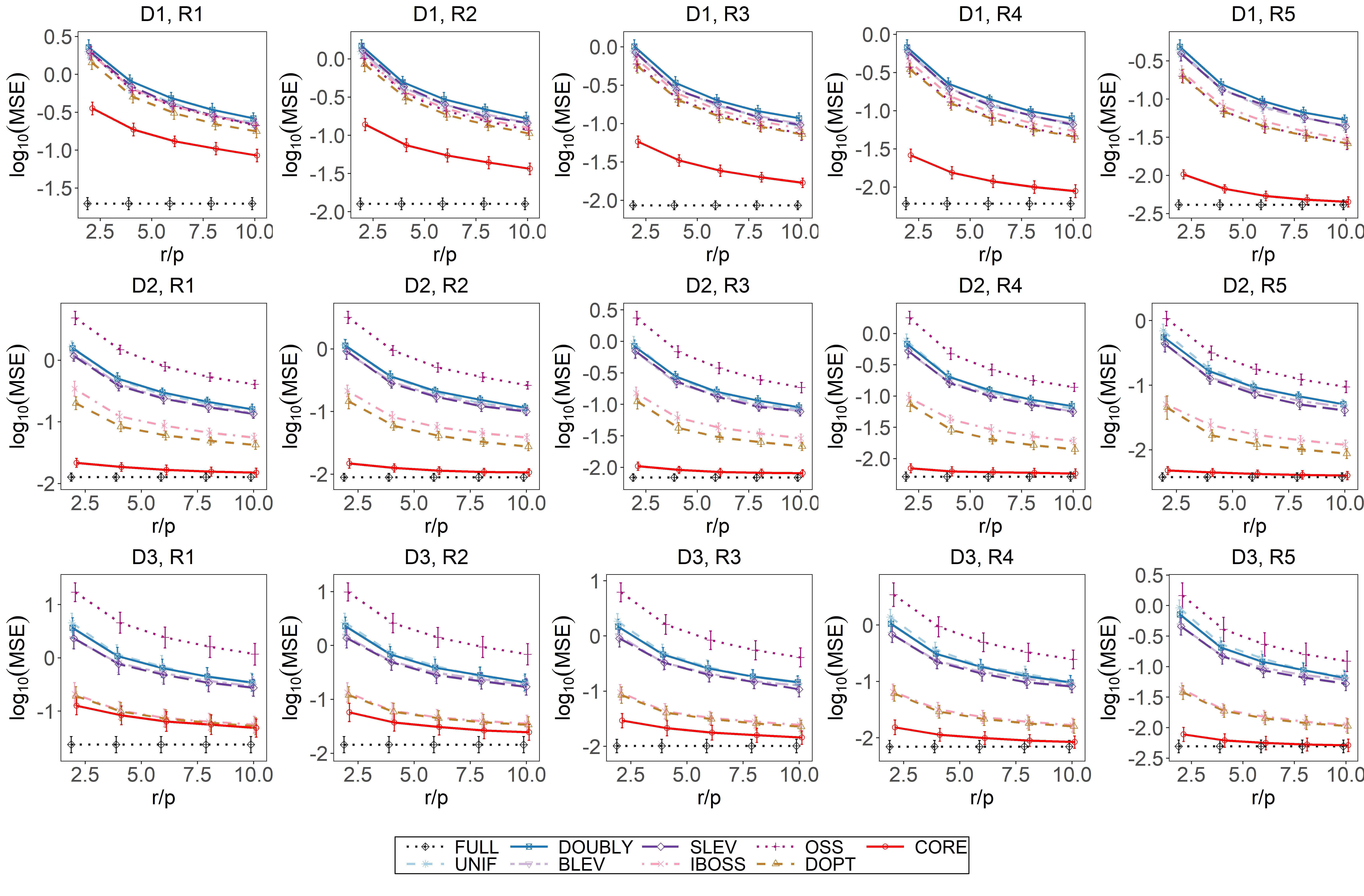}
    \caption{Comparison of different estimators w.r.t. MSE. Each row represents a particular data distribution, i.e., (D1)--(D3), and each column represents a different sparsity ratio, i.e., (R1)--(R5). Vertical bars are the standard errors.}
    \label{fig:simu-mse}
\end{figure}

\begin{figure}[!t]
    \centering
    \includegraphics[width=5in]{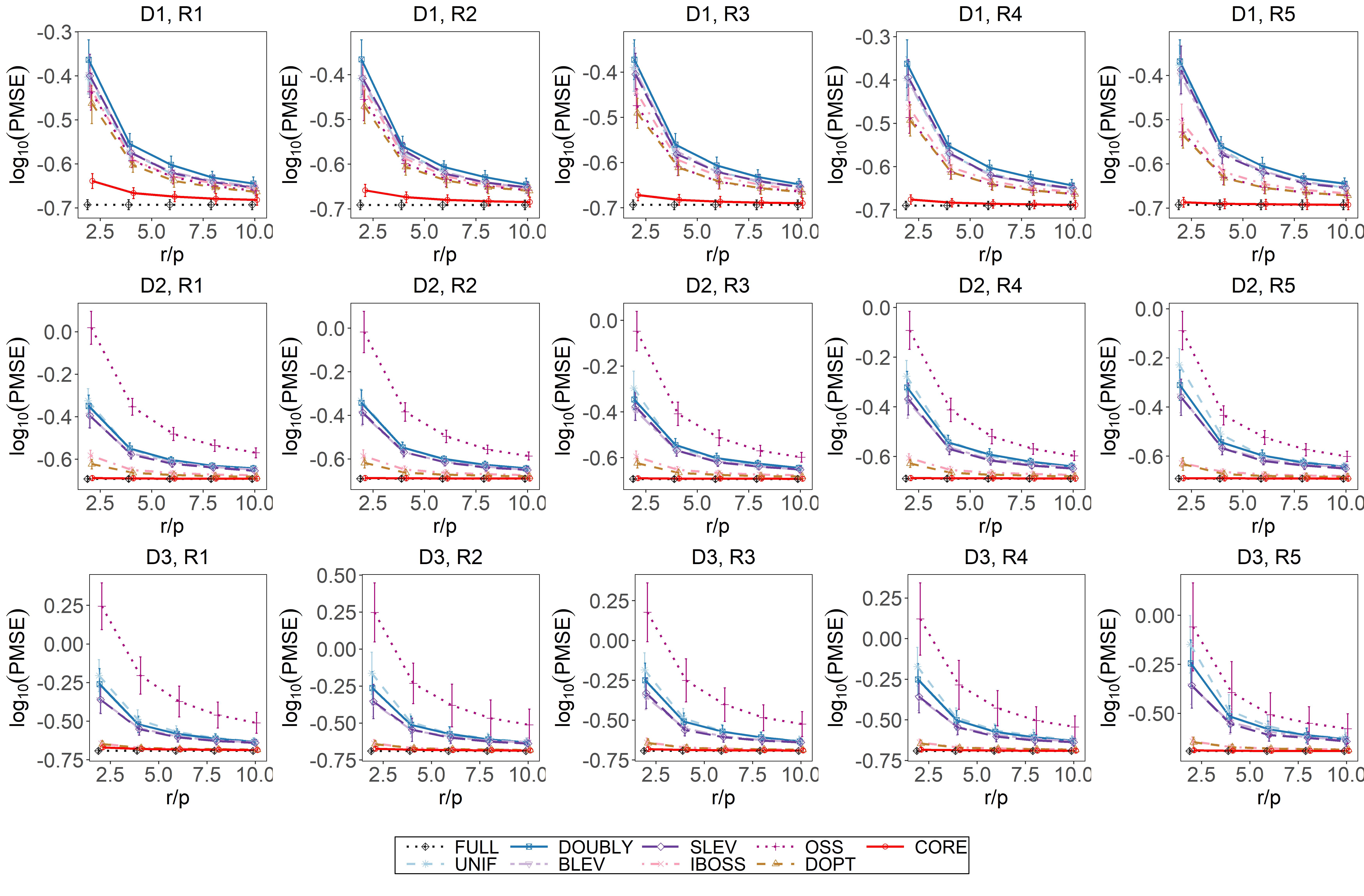}
    \caption{Comparison of different estimators w.r.t. PMSE. Each row represents a particular data distribution, i.e., (D1)--(D3), and each column represents a different sparsity ratio, i.e., (R1)--(R5). Vertical bars are the standard errors.}
    \label{fig:simu-pmse}
\end{figure}

We calculate the empirical mean squared error (MSE) for each of the estimators based on one hundred replications, i.e., 
\begin{equation}\label{eq:mse}
    \operatorname{MSE} = \frac{1}{100}\sum_{i=1}^{100} \frac{\|\widehat{\bbeta}^{(i)}-\bbeta\|^{2}}{\|\bbeta\|^{2}},
\end{equation}
where $\widehat{\bbeta}^{(i)}$ represents the estimator in the $i$th replication.
We also consider the prediction MSE (PMSE).
For the $i$th replication, we randomly split the observed sample into a training set $(\by_{\text{train}}^{(i)}, \X_{\text{train}}^{(i)})$ of size $\lfloor 0.7n \rfloor$ and a test set $(\by_{\text{test}}^{(i)}, \X_{\text{test}}^{(i)})$ of size $\lceil 0.3n \rceil$. 
For each subsampling method, we select a subset from the training set leading to an estimator $\widehat{\bbeta}^{(i)}_{\text{train}}$, and then use it to predict the response $\by_{\text{test}}$ in the test set. 
In this way, PMSE is calculated as
\begin{equation}\label{eq:pmse}
    \operatorname{PMSE} = \frac{1}{100}\sum_{i=1}^{100} \frac{\|\X_{\text{test}}^{(i)}\widehat{\bbeta}^{(i)}_{\text{train}}-\by_{\text{test}}^{(i)}\|^{2}}{\|\by_{\text{test}}^{(i)}\|^{2}}.
\end{equation}
The results of $\log(\operatorname{MSE})$ and $\log(\operatorname{PMSE})$ versus different subsample sizes are shown in Figs.~\ref{fig:simu-mse} and~\ref{fig:simu-pmse}, respectively.

In Figs.~\ref{fig:simu-mse} and~\ref{fig:simu-pmse}, we observe that both MSE and PMSE w.r.t. all estimators decreases as $r$ increases.
We also observe that CORE consistently outperforms other subsampling approaches under all circumstances.
The advantage becomes more apparent when the level of sparsity increases, i.e., from (R1) to (R5).
Such an observation indicates the proposed estimator can provide a more effective estimate than the competitors, especially when the predictor matrix is of high sparsity.
Such success can be attributed to the fact that the proposed core-elements approach can effectively utilize the sparsity structure of the predictor matrix, and the proposed estimator is unbiased and has an approximately minimized estimation variance.

While other deterministic subsampling methods (i.e., IBOSS, OSS, and DOPT) are also competitive, their performance varies significantly across different data distributions. Specifically, the OSS method performs well with normally distributed data (D1), but it loses efficacy with asymmetric log-normal (D2) and heavy-tailed t-distributions (D3), performing worse than uniform sampling. This behavior aligns with the findings presented in Table 3 of \cite{yu2023review}. In contrast, while IBOSS and DOPT are comparable to CORE in handling heavy-tailed distributions, they fall short in normal distributions. Overall, the proposed core-elements approach demonstrates both generality and superiority across a variety of data distributions.

\begin{figure*}[!t]
    \centering
    \includegraphics[width=5in]{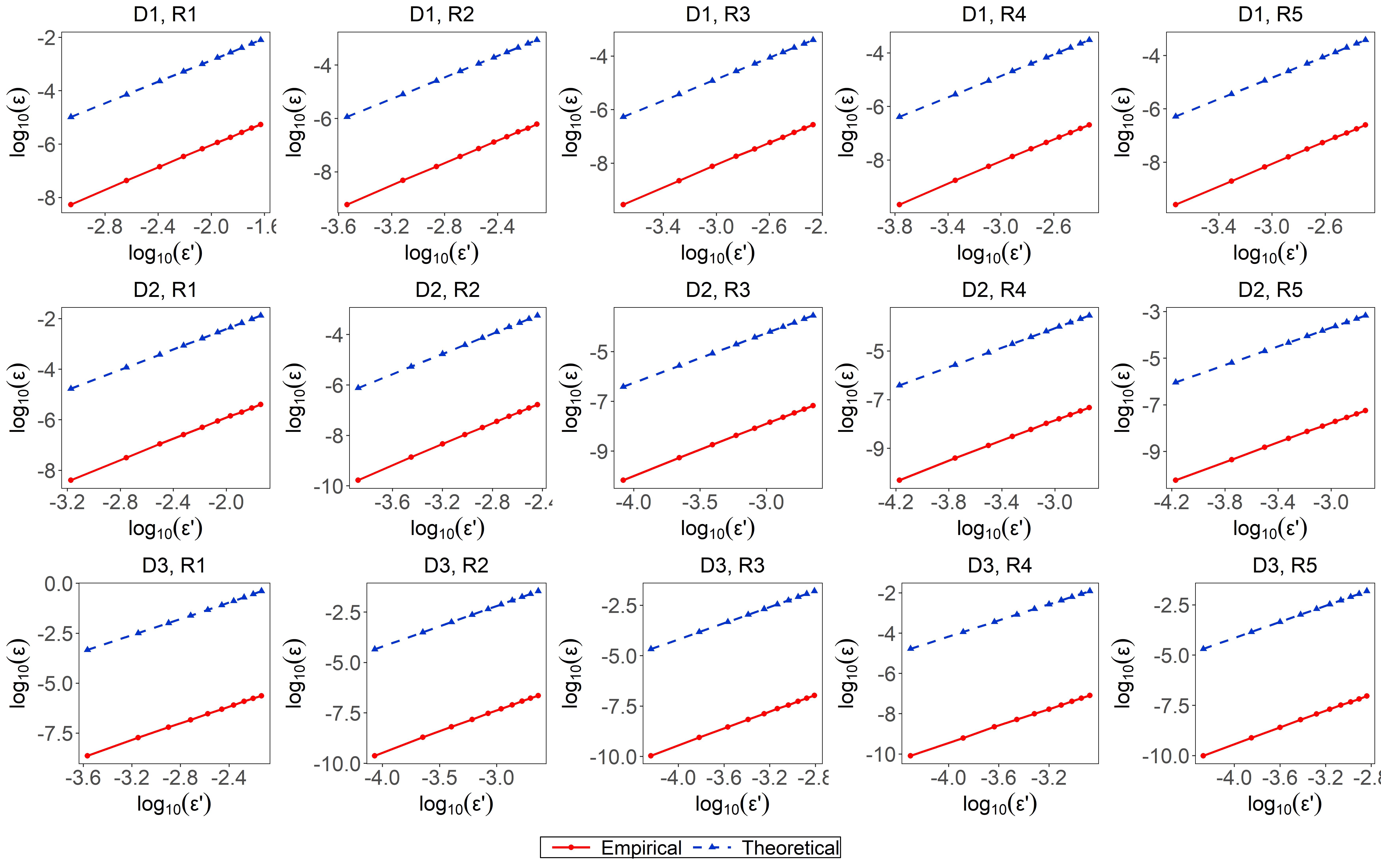}
    \caption{Comparison of the empirical value and the theoretical value of the error term $\epsilon$. Each row represents a particular data distribution, i.e., (D1)--(D3), and each column represents a different sparsity ratio, i.e., (R1)--(R5).}
    \label{fig:eps_error}
\end{figure*}

In order to verify the theoretical error bound provided in Theorem~\ref{th1}, we compare the empirical and theoretical values of $\epsilon$ under different values of $\epsilon^\prime$, as shown in Fig.~\ref{fig:eps_error}. 
Specifically, given a small $\epsilon^\prime$, the empirical value of $\epsilon$ is calculated as
$$\frac{\|\by-\X \widetilde{\bbeta}\|^{2}}{\|\by-\X \hB_{\rm OLS}\|^{2}}-1$$
according to \eqref{eq:th1}, and the theoretical value of $\epsilon$ is calculated as
$$\left[\frac{\epsilon^\prime \kappa^2(\X) \{\kappa^2(\X)+1\}\|\by\|}{\{1-\epsilon^\prime \kappa^2(\X)\}\|\by - \X \hB_{\rm OLS}\|}\right]^2$$
according to \eqref{eq:th1-2}. Both $\epsilon$ and $\epsilon^\prime$ have been made logarithmic transformation in Fig.~\ref{fig:eps_error}.
We can observe that although the empirical and theoretical values of $\epsilon$ differ, their growth trends have an apparent parallel pattern. 
This observation indicates that our proposed error bound and the empirical value are of the same order. Their difference is up to a constant under the log transformation.

\subsection{Performance on Corrupted Data}
We compare the proposed MOM core-elements approach (i.e., Algorithm~\ref{alg:ALG2}), referred to as MOM-CORE, with the full sample OLS estimation and the subsampling methods mentioned above in the presence of outliers. To ensure fairness, all competing methods are equipped with the MOM procedure.

The corrupted data consist of informative data $\{(\bx_{i}, y_{i})\}_{i \in \cI}$ and various types of outliers $\{(\bx_{i}, y_{i})\}_{i \in \cO}$ with $\cO = \cO_1 \cup \cO_2 \cup \cO_3 \cup \cO_4$, such that $|\cO|=n_o$ and $|\cI|=n-n_o$. $\{(\bx_{i}, y_{i})\}_{i \in \cI}$ is generated in the same way as in the previous section, and $\{(\bx_{i}, y_{i})\}_{i \in \cO}$ is constructed following the setup in~\cite{lecue2020robust}. More precisely,
\begin{itemize}
\item for ${i \in \cO_1}$ with $|\cO_1|=\lceil n_o/4 \rceil$, $y_{i}=1000 + 10\zeta_1$ and $\bx_{i}=-10\times 1_p + \zeta_2$, where $\zeta_1\in\RR$ and $\zeta_2\in\RR^p$ are noises following the (multivariate) standard normal distribution;
\item for ${i \in \cO_2}$ with $|\cO_2|=\lceil n_o/4 \rceil$, $y_{i}=-500 + 10\zeta_1$ and $\bx_{i}=10\times 1_p + \zeta_2$;
\item for ${i \in \cO_3}$ with $|\cO_3|=\lceil n_o/4 \rceil$, $y_{i}$ is a $0-1$-Bernoulli random variable and $\bx_{i}$ is uniformly distributed over $[0,1]^{p}$;
\item for ${i \in \cO_4}$, $(\bx_{i}, y_{i})$ is generated from the linear model~\eqref{LM1} with the same true parameter $\bbeta=1_p$ but for a different choice of design $\X$ and noise $\bep$. Here, we take the covariance matrix ${\Sigma}$ as an identity matrix and $\bep$ is a heavy-tailed noise following $t_2$ distribution.
\end{itemize}
After generating the corrupted data above, observations in $\cI$ and $\cO$ are merged and shuffled before downstream operations.
Let the sample size $n = 5 \times 10^4$, the dimension $p = 20$, and the number of outliers $n_o = 19$.
We subsample $r\in \{40p,50p,60p,70p,80p\}$ rows for row-sampling methods or $s=rp$ elements for MOM-CORE. 
The number of blocks is set to be $k=40$ for the MOM procedure. 
Other settings are the same as those in the above section.

\begin{figure}[!t]
    \centering
    \includegraphics[width=5in]{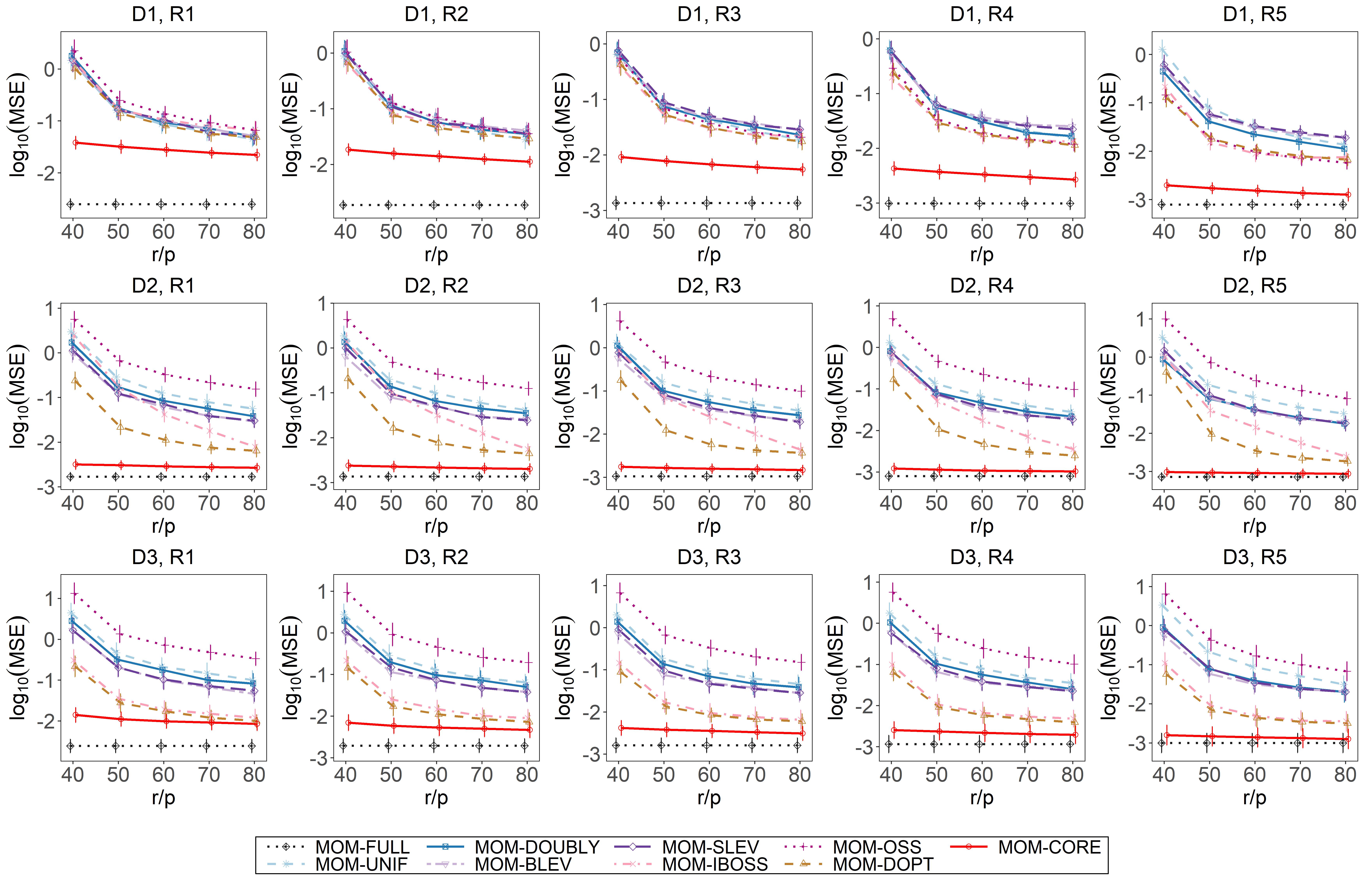}
    \caption{Comparison of different estimators w.r.t. MSE on corrupted data. Each row represents a particular data distribution, i.e., (D1)--(D3), and each column represents a different sparsity ratio, i.e., (R1)--(R5). Vertical bars are the standard errors.}
    \label{fig:simu-mom-mse}
\end{figure}

\begin{figure}[!t]
    \centering
    \includegraphics[width=5in]{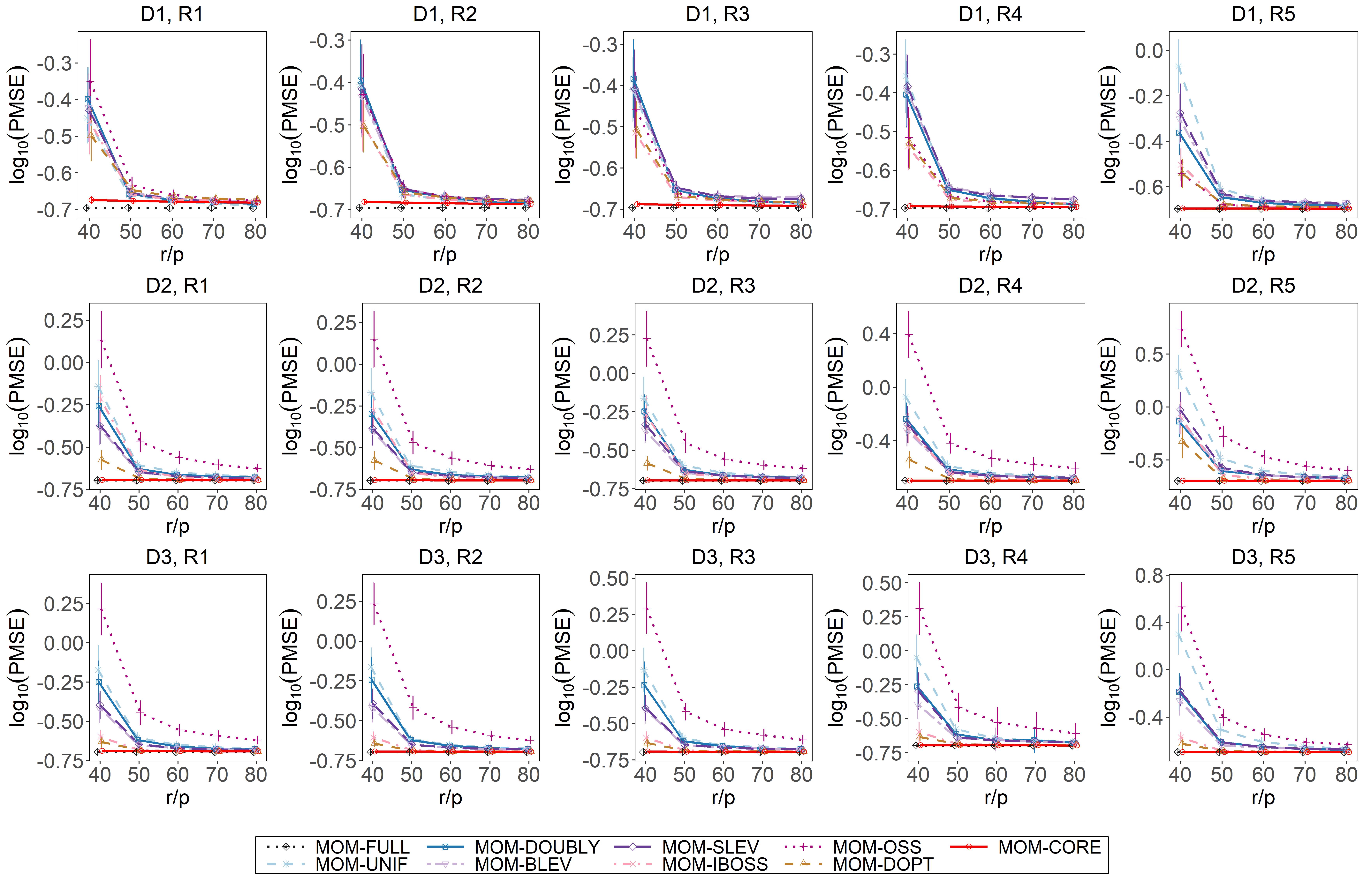}
    \caption{Comparison of different estimators w.r.t. PMSE on corrupted data. Each row represents a particular data distribution, i.e., (D1)--(D3), and each column represents a different sparsity ratio, i.e., (R1)--(R5). Vertical bars are the standard errors.}
    \label{fig:simu-mom-pmse}
\end{figure}

To evaluate the performance of different methods on corrupted data, we calculate MSE according to~\eqref{eq:mse} and PMSE according to~\eqref{eq:pmse} for each method, and their results versus increasing subsample sizes are shown in Figs.~\ref{fig:simu-mom-mse} and~\ref{fig:simu-mom-pmse}, respectively. 
Remark that when predicting the responses, we first split the training set and test set on informative data $\{(\bx_{i}, y_{i})\}_{i \in \cI}$, and then add outliers $\{(\bx_{i}, y_{i})\}_{i \in \cO}$ to the training set; that is, the test set is not corrupted by outliers.

As shown in Figs.~\ref{fig:simu-mom-mse} and~\ref{fig:simu-mom-pmse}, MOM-CORE consistently achieves the smallest estimation and prediction errors among all these subsampling methods, and its advantage becomes more prominent with the increase of sparsity.
Such an observation indicates that by integrating with the MOM procedure, our proposed MOM core-elements algorithm leads to an effective and robust estimator.
Additionally, it is noteworthy that in both Figs.~\ref{fig:simu-pmse} and~\ref{fig:simu-mom-pmse}, the (MOM-)CORE prediction not only outperforms other subsampling approaches but also often delivers results that are nearly indistinguishable from full data.

\subsection{Computing Time}
To compare the computational efficiency of these subsampling approaches, we present the CPU time (in seconds) for different combinations of the sample size $n$ and the dimension $p$ under the case of (D1), (R1) in Table~\ref{tab:simu-time}. 
Here, data corruption and the MOM procedure are omitted to save space, as they hardly affect the computation time.
We take the subsample parameter $r=10p$.
All computations are implemented using the R programming language on a desktop running Windows 10 with an Intel i5-10210U CPU and 16GB memory.
The CPU time for using the full sample is also presented for comparison.

\begin{table}[!t]
\centering
\caption{CPU time (in seconds) of estimating $\bbeta$, for different combinations of $n$ and $p$ under (D1), (R1).}
\label{tab:simu-time}
\resizebox{\textwidth}{!}{
\begin{tabular}{lrrrrrrrrr}
\multicolumn{10}{c}{(a) CPU time for different $p$, with fixed $n=10^{5}$.}\\
\toprule
Method & FULL & UNIF & DOUBLY & BLEV & SLEV & IBOSS & OSS & DOPT & CORE \\
\midrule
$p=50$ & 0.82 & 0.00 & 0.02 & 0.51 & 0.51 & 0.15 & 0.40 & 0.43 & 0.08 \\
$p=100$ & 3.19 & 0.01 & 0.08 & 2.02 & 2.03 & 0.43 & 2.26 & 1.89 & 0.35 \\
$p=500$ & 76.31 & 0.39 & 1.61 & 46.37 & 46.39 & 6.53 & 58.77 & 45.54 & 5.29 \\
\bottomrule
\multicolumn{10}{c}{(b) CPU time for different $n$, with fixed $p=100$.}\\
\toprule
Method & FULL & UNIF & DOUBLY & BLEV & SLEV & IBOSS & OSS & DOPT & CORE \\
\midrule 
$n=5\times10^4$ & 1.72 & 0.00 & 0.06 & 1.01 & 1.01 & 0.12 & 1.87 & 0.89 & 0.13 \\
$n=5\times10^5$ & 16.95 & 0.02 & 0.10 & 10.86 & 10.85 & 1.36 & 6.41 & 9.92 & 0.87 \\
$n=5\times10^6$ & 189.65 & 0.11 & 0.21 & 119.83 & 119.84 & 12.94 & 43.28 & 91.52 & 7.54 \\
\bottomrule
\end{tabular}
}
\end{table}

As can be observed from Table~\ref{tab:simu-time}, all of these estimates are more efficient than the full sample OLS estimate.
It is unsurprising that the random sampling-based methods, UNIF and DOUBLY, require the least computing time, as they avoid the additional step of calculating subsampling probabilities. Among these methods, BLEV, SLEV, and DOPT necessitate the calculation of singular value decomposition of the full predictor matrix, leading to relatively longer CPU times. As expected, OSS also requires considerable computing time due to its complexity of $O(np \log r + rp^2)$.

Notably, the proposed core-elements approach only requires a longer time than UNIF and DOUBLY, while being faster than other competitors in almost all circumstances. In addition, its superiority in computation becomes more prominent when $n$ is much larger than $p$, which is exactly the most suitable situation for taking advantage of subsampling.

\section{Real Data Example}\label{sec:real}

The rapid development of the single-cell RNA sequencing (scRNA-seq) technique enables the gene expression profiling of single cells. 
ScRNA-seq data are often organized into a reads count matrix, where rows are cells, columns represent genes, and the $(i,j)$th component is the observed expression level of the gene $j$ in cell $i$.
We consider a scRNA-seq dataset collected by \cite{azizi2018single}, which includes $\operatorname{CD}45+$ immune cells from eight breast carcinomas, as well as matched normal breast tissue, blood, and lymph node. The dataset is publicly available with the accession code GSE114725 in Gene Expression Omnibus \citep{edgar2002gene}.
Our goal is to find the relationship between the expression of the gene MT-RNR2 and other genes. 
As a critical neuroprotective factor, the MT-RNR2 gene encodes the Humanin polypeptide and protects against death in Alzheimer's disease.

\begin{figure}[!t]
    \centering
    \subfigure[Box plots of predictor leverage values (left) and response values (right).]{\includegraphics[height=1.2in]{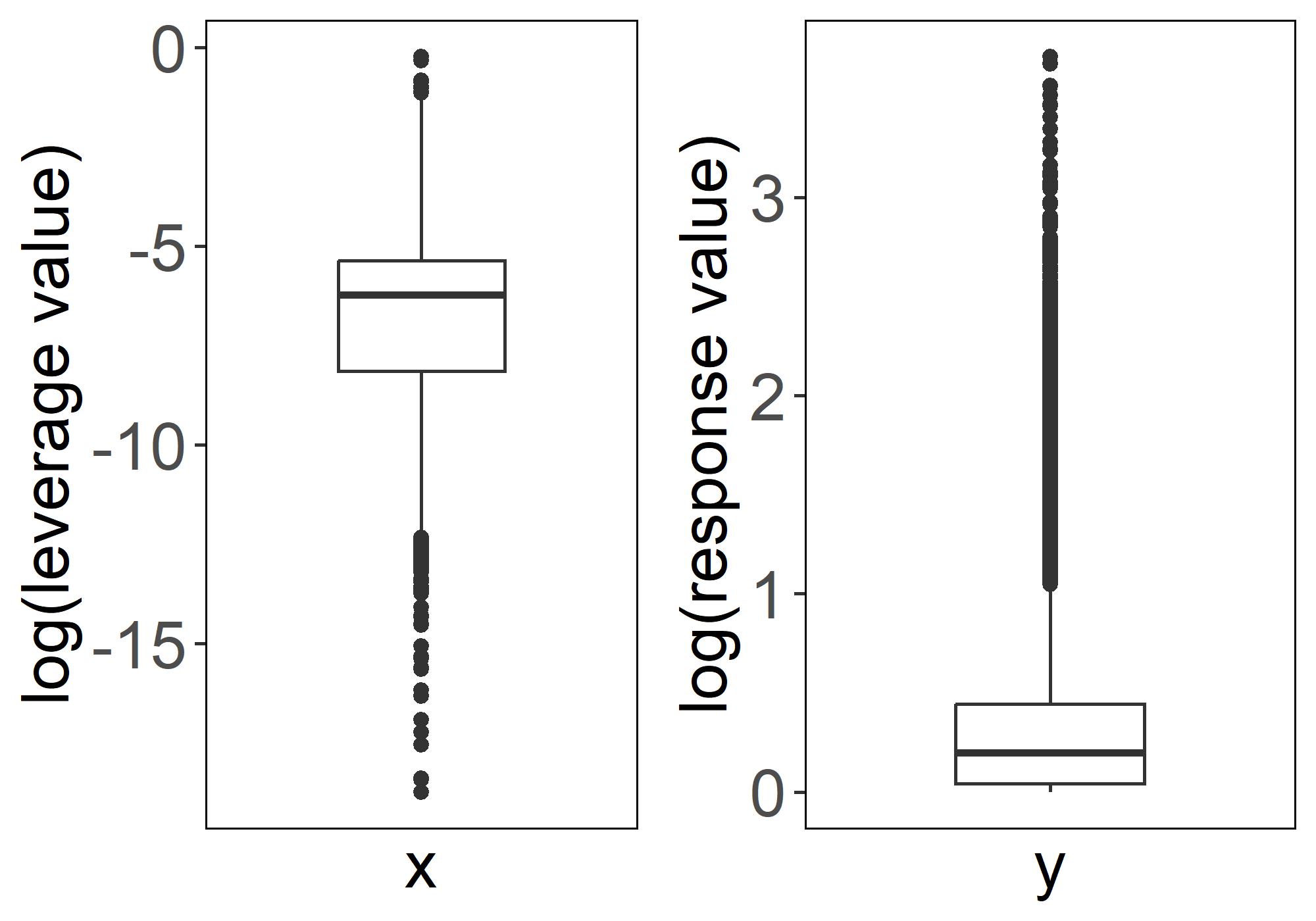}
    \label{fig:scRNA-a}}
    \hspace{0.3cm}
    \subfigure[Histogram of values in the predictor matrix.]{\includegraphics[height=1.2in]{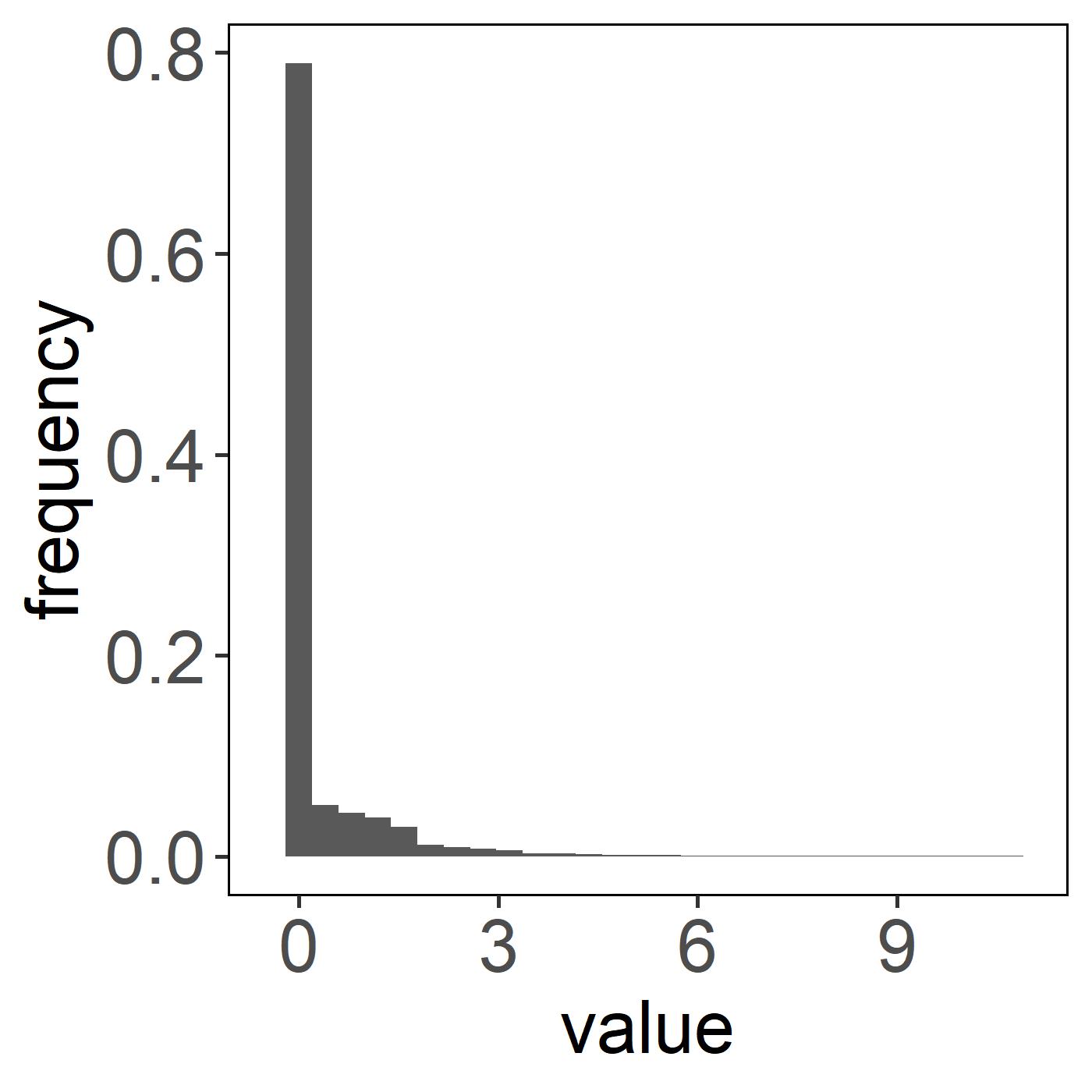}
    \label{fig:scRNA-b}}
    \subfigure[Hexbin scatter plots~\citep{carr2023hexbin} of the response against ten randomly selected predictors.]{\includegraphics[width=3.4in]{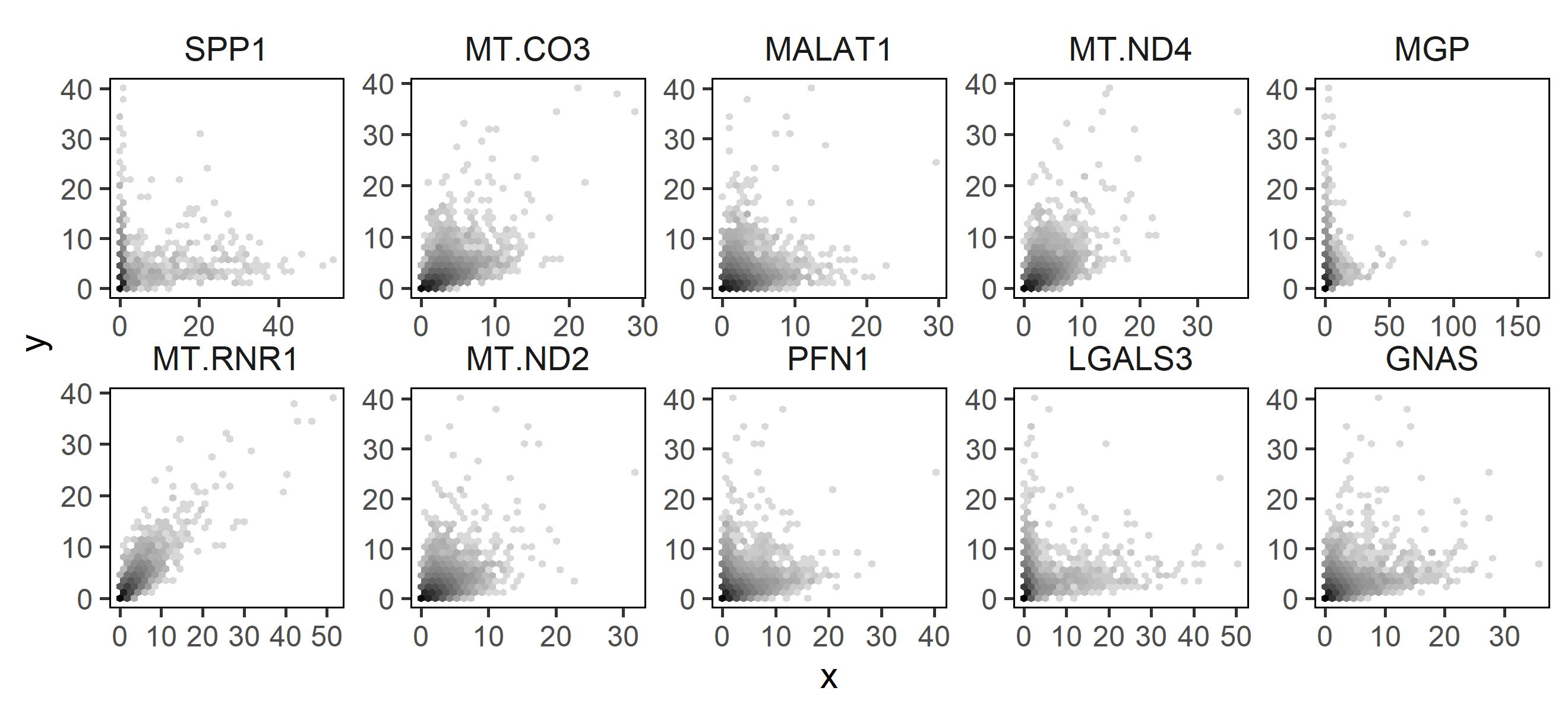}
    \label{fig:scRNA-c}}
    \caption{Visualization of the scRNA-seq dataset's distribution. For visual clarity, the leverage and response values are (shifted to the positive range and) log-transformed in the subfigure (a), and the predictor values are truncated at the 99.9\% quantile in the subfigure (b).}
    \label{fig:scRNA-vis}
\end{figure}

To achieve the goal, we take the reads of this gene as the response and the reads for other genes as predictors. 
Following the data pre-processing steps in \cite{huang2008adaptive}, we first screen the genes as follows: (1) select the top $3000$ genes with the highest expression levels; and (2) select the top $500$ genes with the largest variances.
We then standardize the predictors so that they have unit variance.
The distribution of pre-processed data points is shown in Fig.~\ref{fig:scRNA-a}.
The final sample contains $n \approx 10^5$ cells and $p = 500$ genes with over 75\% zero elements, as illustrated through the histogram in Fig.~\ref{fig:scRNA-b}. 
Figure~\ref{fig:scRNA-c} presents the relationship between the response against ten randomly selected predictors of the scRNA-seq dataset, from which we can observe linear patterns. Therefore, it is reasonable to assume the data follow the linear model~\eqref{LM1}.

As the true coefficient vector $\bbeta$ is unknown in real-world data analysis, we focus on comparing prediction performance by calculating the PMSE as defined in~\eqref{eq:pmse}. This evaluation is based on one hundred bootstrap samples. The training and test sets are randomly partitioned according to the ratio of 7:3. The subsampling methods considered here are the same as those in Section~\ref{sec:simu}, and the subsample size is set to be $r\in\{2p, 2^2p, 2^3p, 2^4p, 2^5p\}$ rows or $s = rp$ elements equivalently. We set the number of blocks for MOM to be $k=5$.

As shown in Fig.~\ref{fig:scRNA-vis}, the data exhibits linear patterns between the response and predictors, with no apparent outliers. Consequently, the MOM variant performs similarly to its original estimator. For clarity, we present only the MOM-FULL and MOM-CORE, omitting the MOM variants of other competing subsampling methods.

\begin{figure}[!t]
    \centering
    \includegraphics[height=1.3in]{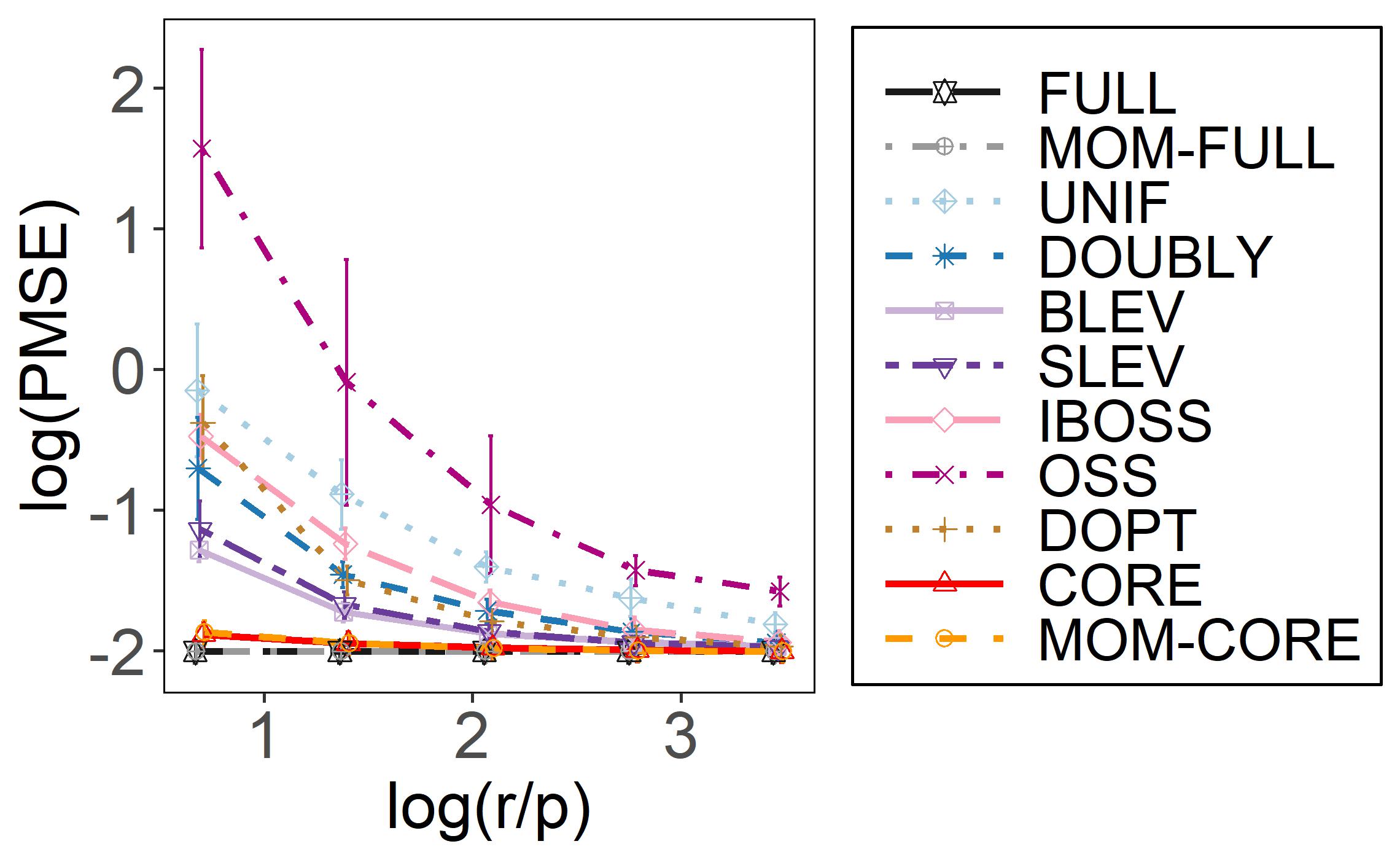}
    \caption{Comparison of different methods w.r.t. PMSE for the scRNA-seq dataset.}
    \label{fig:scRNA-d}
\end{figure}

Figure~\ref{fig:scRNA-d} displays the performance of different approaches for predicting $\by$ on the test set. 
We observe the baseline methods, FULL and MOM-FULL, achieve almost the same prediction accuracy, suggesting the absence of extreme outliers in the dataset.
Similarly, the proposed CORE and MOM-CORE methods have nearly identical prediction accuracy, both of which consistently outperform other subsampling approaches.
In addition, the proposed estimators perform almost the same as $\hB_{\rm OLS}$ w.r.t. the PMSE, even when the selected number of elements is just $s=2p^2$.

Table~\ref{tab:scRNA-time} shows the CPU time for these subsampling methods.
Compared to the full sample estimate, all the subsample estimates require a shorter computational time, except for the OSS method as $r$ increases. This exception can be attributed to the multiplicative order $O(\log r)$ in its computational cost.
Similar to the results in Table~\ref{tab:simu-time}, we observe that the core-elements approach has great advantages in computing. It is only slower than the random sampling-based UNIF and DOUBLY methods, while requiring nearly half the CPU time of the IBOSS method.
The results in Fig.~\ref{fig:scRNA-d} and Table~\ref{tab:scRNA-time} indicate that the proposed strategy can provide a more effective estimate than the competitors, requiring a relatively short computational time.

\begin{table}[!t]
\centering
\caption{CPU time (in seconds) of estimating $\bbeta$ for the scRNA-seq dataset.}
\label{tab:scRNA-time}
\resizebox{\textwidth}{!}{
\begin{tabular}{lrrrrrrrrr}
\toprule
Method & FULL & UNIF & DOUBLY & BLEV & SLEV & IBOSS & OSS & DOPT & CORE \\
\midrule
$r=2p$ & - & 0.19 & 0.23 & 46.98 & 46.96 & 8.27 & 12.13 & 43.29 & 3.24 \\
$r=2^3p$ & - & 0.81 & 1.16 & 47.44 & 47.43 & 9.81 & 85.32 & 44.84 & 6.05 \\
$r=2^5p$ & - & 3.39 & 6.43 & 50.06 & 50.08 & 15.33 & 438.91 & 47.50 & 8.53 \\
$n$ & 80.24 & - & - & - & - & - & - & - & - \\
\bottomrule
\end{tabular}
}
\end{table}

\section{Discussion}\label{sec:conclude}
Realizing the gaps of element-wise subset selection methods in large-scale data analysis, we developed the core-elements method for least squares estimation in linear models.
Theoretically, we showed that the proposed core-elements estimator approximately minimizes an upper bound of the estimation variance.
We also provided a coresets-like finite-sample bound for the proposed estimator, which is supported by empirical results.
To deal with data corruption, we introduced the median-of-means estimation to provide a robust version of core-elements and established consistency of the resultant estimator.
Empirical studies suggest that the proposed method is not only suitable for (numerically) sparse matrices but also has a superior performance for more general dense cases in terms of both accuracy and time.

Considering that the predictor matrix defined by basis function evaluations usually enjoys the (numerically) sparse property, we plan to extend core-elements to nonparametric additive models for efficiently approximating the penalized least squares estimation in the future.
More importantly, the core-elements approach not only facilitates computation, but also has excellent applications in preserving data privacy and improving communication efficiency in federated learning, which are also left to our future work.

\bigskip
\begin{center}
{\large\bf SUPPLEMENTARY MATERIAL}
\end{center}
The Supplementary Material is structured as follows. Section~\ref{sec:proof} provides technical details of the theoretical results stated within the manuscript. The additional numerical results in Section~\ref{sec:numerical} evaluate the performance of the proposed method in misspecified linear models and other choices of model coefficients.

\section{Technical details}\label{sec:proof}

\subsection{Proof of Lemma~\ref{lem3}}
\begin{proof}
The variance of $\tB$ can be organized as
\begin{align}
    E(\|\tB-\bbeta\|^{2}|\X) =& E[\{(\X^{\ast \top} \X)^{-1} \X^{\ast \top} \by-\bbeta\}^\top\{(\X^{\ast \top} \X)^{-1} \X^{\ast \top} \by-\bbeta\}] \nonumber\\
    =& E\{\by^\top \X^{\ast} (\X^{\top} \X^\ast)^{-1} (\X^{\ast \top} \X)^{-1} \X^{\ast \top} \by\} -\bbeta^\top \bbeta \nonumber\\
    =& \sigma^2\|(\X^{\ast \top} \X)^{-1} \X^{\ast \top}\|_F^2. \label{eq:proof-lem3-1}
\end{align}
The last equality is due to 
$$E(\by^\top A \by) = (\X\bbeta)^\top A \X\bbeta + \sigma^2 \tr(A)$$
with $$A=\X^{\ast} (\X^{\top} \X^\ast)^{-1} (\X^{\ast \top} \X)^{-1} \X^{\ast \top}.$$ 

Recalling that $\X^\ast = \X-\bL$, the term $\|(\X^{\ast \top} \X)^{-1} \X^{\ast \top}\|_F^2$ in \eqref{eq:proof-lem3-1} equals
\begin{align*}
& \tr\{(\X-\bL) (\X^{\top}\X-\X^\top\bL)^{-1} (\X^{\top}\X-\bL^\top\X)^{-1} (\X-\bL)^{\top}\} \\
=& \tr[(\X-\bL) (\X^{\top}\X)^{-1} \{\I_p-\X^{\top}\bL(\X^{\top}\X)^{-1}\}^{-1} \{\I_p-(\X^{\top}\X)^{-1}\bL^{\top}\X\}^{-1} (\X^{\top}\X)^{-1} (\X-\bL)^{\top} ] \\
=& \tr[\{\I_p-\X^{\top}\bL(\X^{\top}\X)^{-1}\}^{-1} \{\I_p-(\X^{\top}\X)^{-1}\bL^{\top}\X\}^{-1} (\X^{\top}\X)^{-1} (\X-\bL)^{\top} (\X-\bL)(\X^{\top}\X)^{-1}] \\
=& \tr(\bS_1 \bS_2),
\end{align*}
where 
$$\bS_1 =\{\I_p-\X^{\top}\bL(\X^{\top}\X)^{-1}\}^{-1} \{\I_p-(\X^{\top}\X)^{-1}\bL^{\top}\X\}^{-1}$$ 
and 
$$\bS_2 = (\X^{\top}\X)^{-1} (\X-\bL)^{\top} (\X-\bL)(\X^{\top}\X)^{-1}.$$
Considering that $\bS_2$ is a positive semi-definite (PSD) matrix, it holds that
\begin{equation}\label{eq:lem2-3}
    \tr(\bS_1 \bS_2) \leq \sum_i \sigma_i(\bS_1) \sigma_i(\bS_2) \leq \sum_i\|\bS_1\|_2 \sigma_i(\bS_2)=\sum_i\|\bS_1\|_2 \lambda_i(\bS_2)=\|\bS_1\|_2 \tr(\bS_2),
\end{equation}
where $\sigma_i(\cdot)$ and $\lambda_i(\cdot)$ stand for the $i$th singular value and eigenvalue, respectively; the first inequality comes from the Von Neumann's trace inequality, and the last-but-one equality holds because the matrix $\bS_2$ is PSD. Therefore, it suffices to bound the terms $\|\bS_1\|_2$ and $\tr(\bS_2)$.

First, we consider the term $\|\bS_1\|_2$. Performing a Taylor expansion of $\{\I_p-(\X^{\top}\X)^{-1}\bL^{\top}\X\}^{-1}$ around the point $\bL = 0_{n \times p}$ yields
$$
\{\I_p-(\X^{\top}\X)^{-1}\bL^{\top}\X\}^{-1} = \I_p+(\X^{\top}\X)^{-1}\bL^{\top}\X+\bW_1
$$
under the convergence condition that the spectral radius $\|(\X^{\top}\X)^{-1}\bL^{\top}\X\|_2 = \lambda_0 <1$; see \citet[Chap.~1]{higham2008functions}. Then, the remainder satisfies $\|\bW_1\|_2=O(\lambda_0^2)$.
Based on the result of Taylor expansion, we have
\begin{align*}
\bS_1 =& \{\I_p+\X^{\top}\bL(\X^{\top}\X)^{-1}+\bW_1^\top\} \{\I_p+(\X^{\top}\X)^{-1}\bL^{\top}\X+\bW_1\} \\
=& \I_p+\X^{\top}\bL(\X^{\top}\X)^{-1}+(\X^{\top}\X)^{-1}\bL^{\top}\X+\bW_2
\end{align*}
with
\begin{align*}
\bW_2 = &\bW_1 + \bW_1^\top + \bW_1^\top \bW_1 + \bW_1^\top (\X^{\top}\X)^{-1}\bL^{\top}\X \\
&+ \X^{\top}\bL (\X^{\top}\X)^{-1} \bW_1 + \X^{\top}\bL(\X^{\top}\X)^{-2}\bL^{\top}\X.
\end{align*}
One can see that $\|\bW_2\|_2 = O(\lambda_0^2)$. Consequently, we have 
\begin{equation}\label{eq:lem2-1}
    \|\bS_1\|_2 \le 1 + O(\lambda_0).
\end{equation}

Next, we consider the other term, $\tr(\bS_2)$. It holds that
\begin{align}\label{eq:lem1-trS2}
\tr(\bS_2) = \tr\{(\X^{\top}\X )^{-1}\} - 2\tr\{(\X^{\top}\X)^{-1}\bL^{\top}\X(\X^{\top}\X )^{-1}\} + \tr\{(\X^{\top}\X)^{-2}\bL^{\top}\bL\}.
\end{align}
As both $(\X^{\top}\X)^{-1}$ and $\bL^\top \bL$ are PSD matrices, we apply the inequality~\eqref{eq:lem2-3} on~\eqref{eq:lem1-trS2} and obtain that
\begin{align}
    \tr(\bS_2) &\le \tr\{(\X^{\top}\X )^{-1}\} + 2\|(\X^{\top}\X)^{-1}\bL^{\top}\X\|_2 \tr\{(\X^{\top}\X )^{-1}\} + \|(\X^{\top}\X)^{-2}\|_2\tr(\bL^{\top}\bL) \nonumber \\ 
    &\le \tr\{(\X^{\top}\X )^{-1}\} + 2\lambda_0 \tr\{(\X^{\top}\X )^{-1}\} + \|(\X^{\top}\X)^{-1}\|_2^2 \|\bL\|_F^2. \label{eq:lem2-2}
\end{align}

By combining \eqref{eq:lem2-1} and \eqref{eq:lem2-2}, we conclude that
\begin{align*}
\tr(\bS_1 \bS_2) \le& \|\bS_1\|_2 \tr(\bS_2) \\
\le& [\tr\{(\X^{\top}\X)^{-1}\} + \|(\X^{\top}\X)^{-1}\|_2^2 \|\bL\|_F^2]\{1+O(\lambda_0)\}.
\end{align*}
Therefore, the variance of $\tB$ can be bounded by
\begin{align*}
E(\|\tB-\bbeta\|^{2}|\X) \le \sigma^2 [\tr\{(\X^{\top}\X)^{-1}\} + \|(\X^{\top}\X)^{-1}\|_2^2 \|\bL\|_F^2]\{1+O(\lambda_0)\}.
\end{align*}
\end{proof}

\subsection{Proof of Theorem~\ref{th1}}
\begin{proof}
(1) Recalling the definition of OLS estimation, i.e.,
$$\hB_{\rm OLS} = \arg \min _{\btheta \in \RR^{p}} \|\by - \X \btheta\|^{2},$$
the left-hand side of the inequality is obviously established.

(2) For the right-hand side of inequality,
\begin{align*}
\|\by - \X \widetilde{\bbeta}\|^2 =& \|\by - \X \hB_{\rm OLS}\|^2 + 2(\by - \X \hB_{\rm OLS})^\top (\X \hB_{\rm OLS} - \X \widetilde{\bbeta}) + \|\X \hB_{\rm OLS} - \X \widetilde{\bbeta}\|^2 \\
=& \|\by - \X \hB_{\rm OLS}\|^2 + \|\X \hB_{\rm OLS} - \X \widetilde{\bbeta}\|^2.
\end{align*}
The last equality holds because the cross term $(\by - \X \hB_{\rm OLS})^\top (\X \hB_{\rm OLS} - \X \widetilde{\bbeta})$ equals to 0.
Thus, it is sufficient to show that
\begin{equation}\label{eq:pr1}
\|\X \hB_{\rm OLS} - \X \widetilde{\bbeta}\|^2 \le \epsilon \|\by - \X \hB_{\rm OLS}\|^2.
\end{equation}
Simple algebra yields that
\begin{align}
\|\X \hB_{\rm OLS} - \X \widetilde{\bbeta}\|^2
=& \|\X(\X^\top \X)^{-1}\X^\top \by - \X(\X^{\ast\top} \X)^{-1}\X^{\ast\top} \by\|^2 \nonumber\\
\le& \|\X(\X^\top \X)^{-1}\X^\top - \X(\X^{\ast\top} \X)^{-1}\X^{\ast\top}\|_2^2 \|\by\|^2. \label{eq:pr3}
\end{align}
Then, we consider the term $\|\X(\X^\top \X)^{-1}\X^\top - \X(\X^{\ast\top} \X)^{-1}\X^{\ast\top}\|_2^2$. Recall that $\bL = \X-\X^\ast$. By using the condition $\|\bL\|_{2}<\epsilon^\prime\|\X\|_{2}$, we have
\begin{align}
& \|\X(\X^\top \X)^{-1}\X^\top - \X(\X^{\ast\top} \X)^{-1}\X^{\ast\top}\|_2 \nonumber \\
=& \|\X(\X^\top \X)^{-1}\X^\top - \X\{(\X-\bL)^{\top} \X\}^{-1}(\X-\bL)^{\top}\|_2 \nonumber \\
=& \|\X\{(\X^\top \X)^{-1}-(\X^\top\X-\bL^\top\X)^{-1}\}\X^\top + \X(\X^\top\X-\bL^\top\X)^{-1}\bL^\top\|_2 \nonumber\\
\le& \|\X\|_2^2 \|(\X^\top \X)^{-1}-(\X^\top\X-\bL^\top\X)^{-1}\|_2 + \epsilon^\prime \|\X\|_2^2 \|(\X^\top\X-\bL^\top\X)^{-1}\|_2 \nonumber \\
\le& \lambda_0 \|\X\|_2^2 \|(\X^\top\X-\bL^\top\X)^{-1}\|_2 + \epsilon^\prime \|\X\|_2^2 \|(\X^\top\X-\bL^\top\X)^{-1}\|_2 \nonumber \\
=& (\lambda_0 + \epsilon^\prime) \|\X\|_2^2 \|(\X^\top\X-\bL^\top\X)^{-1}\|_2, \label{eq:pr4}
\end{align}
where $\lambda_0 = \|(\X^\top \X)^{-1}\bL^\top \X\|_2$.
The last inequality is due to $A^{-1}-\bB^{-1} = A^{-1}(\bB-A)\bB^{-1}$ and then $\|A^{-1}-\bB^{-1}\|_2 \le \|A^{-1}(\bB-A)\|_2 \|\bB^{-1}\|_2$. 

Consider the term $\|(\X^\top\X-\bL^\top\X)^{-1}\|_2$ in \eqref{eq:pr4}.
A special case of Woodbury formula \citep{hager1989updating} takes the form
$(A-\bB)^{-1}=A^{-1}+A^{-1} \bB(A-\bB)^{-1}$,
which has a recursive structure that yields
\begin{equation}\label{eq:pr5}
    (A-\bB)^{-1}=\sum_{k=0}^{\infty}(A^{-1} \bB)^{k} A^{-1}.
\end{equation}
By using \eqref{eq:pr5}, it follows that
$$
(\X^\top\X-\bL^\top\X)^{-1} = \sum_{k=0}^{\infty} \{(\X^\top\X)^{-1}\bL^\top\X\}^{k} (\X^\top\X)^{-1}.
$$
Thus, we have
\begin{align*}
    \|(\X^\top\X-\bL^\top\X)^{-1}\|_2 \le& \sum_{k=0}^{\infty} \|(\X^\top\X)^{-1}\bL^\top\X\|_2^{k} \|(\X^\top\X)^{-1}\|_2 \\
    =& \frac{1}{1-\lambda_0} \|(\X^\top\X)^{-1}\|_2.
\end{align*}
In this way, the formula \eqref{eq:pr4} can be further bounded by
\begin{align}
\|\X(\X^\top \X)^{-1}\X^\top - \X(\X^{\ast\top} \X)^{-1}\X^{\ast\top}\|_2 &\le \frac{\lambda_0 + \epsilon^\prime}{1-\lambda_0} \|\X\|_2^2 \|(\X^\top\X)^{-1}\|_2 \nonumber \\
&= \frac{\lambda_0 + \epsilon^\prime}{1-\lambda_0} \kappa^2(\X), \label{eq:pr6}
\end{align}
where $\kappa(\X) = \sigma_{\max}(\X)/\sigma_{\min}(\X)$ is the condition number of $\X$.

Considering the spectral radius $\lambda_0$, we have
\begin{align}\label{eq:pr7}
\lambda_0 = \|(\X^\top\X)^{-1}\bL^\top\X\|_2 \leq \epsilon^{\prime} \|(\X^\top\X)^{-1}\|_2 \|\X\|_2^2 = \epsilon^{\prime} \kappa^2(\X).
\end{align}
Combining \eqref{eq:pr3}, \eqref{eq:pr6} and \eqref{eq:pr7}, we conclude that if $\epsilon^\prime$ satisfies
\begin{equation}\label{eq:pr8}
\frac{\{\kappa^2(\X)+1\}\epsilon^\prime}{1-\epsilon^\prime\kappa^2(\X)} \kappa^2(\X) \le \frac{\epsilon^{1/2} \|\by - \X \hB_{\rm OLS}\|}{\|\by\|}
\end{equation}
and
\begin{equation}\label{eq:pr9}
\epsilon^\prime < \frac{1}{\kappa^2(\X)},
\end{equation}
then the desired inequality \eqref{eq:pr1} holds. The inequality \eqref{eq:pr9} is required to ensure the upper bound \eqref{eq:pr7} of $\lambda_0$ is smaller than 1. Further, the inequality \eqref{eq:pr8} can be rewritten as
\begin{equation*}
\epsilon^\prime \le \frac{1}{\kappa^2(\X)} \left[1+\frac{\{\kappa^2(\X)+1\}\|\by\|}{\epsilon^{1/2} \|\by - \X \hB_{\rm OLS}\|}\right]^{-1} < \frac{1}{\kappa^2(\X)}.
\end{equation*}

This completes the proof of the right-hand side of the inequality.
\end{proof}

\subsection{Proof of Remark~\ref{remark1}}
\begin{proof}
Recall that $F$ is the cumulative distribution function of non-zero elements in $\X$. Further assume the distribution is symmetric about zero, which is satisfied under the two specific cases we are concerned with. 

We first consider the general continuous and symmetric $F$.
Let $n_\alpha = \lfloor \alpha n \rfloor$.
For $i=1, \ldots, n_\alpha$ and $j=1, \ldots, p$, define $z_{(i) j}$ be the $i$th order statistic for non-zero elements in $\{x_{1 j}^2, \ldots, x_{n j}^2\}$.
As before, let $\bL = \X-\X^{*}$. Then, we have 
$$
\|\bL\|_{2}^2\leq \|\bL\|_{F}^2 = \sum_{i=1}^{n_\alpha - r}\sum_{j=1}^{p} z_{(i) j} \leq (n_\alpha-r)\sum_{j=1}^{p} z_{(n_\alpha-r) j}.
$$
Define $G(x) = F(\sqrt{x}) - F(-\sqrt{x})$ and $g(x) = \mathrm{d} G(x)/\mathrm{d} x$ for $x\geq 0$ as the cumulative distribution function and probability density function of non-zero elements in $\{x_{1 j}^2, \ldots, x_{n j}^2\}$, respectively. By using the central limit theorem, $z_{(i) j}$ is asymptotically normal with mean $\mu_{i} = G^{-1}(i/n_\alpha)$ and variance $\sigma_{i}^2 = i(n_\alpha-i)/\{n_\alpha^3 g^2(\mu_i)\}$~\citep{walker1968note}. Thus, when $n_\alpha \to \infty$, we have the upper deviation inequality for $j=1, \ldots, p$,
$$
\pr\left(z_{(n_\alpha-r) j} \geq \mu_{n_\alpha-r} + t\right) \leq \exp\left(-\frac{t^2}{2\sigma_{n_\alpha-r}^2}\right)
\quad \text{for any } t \geq 0,
$$ 
that is,
$$
\pr\left[z_{(n_\alpha-r) j} < \mu_{n_\alpha-r} + \{-2\log(\delta)\}^{1/2} \sigma_{n_\alpha-r}\right] > 1-\delta
\quad \text{for any } 0<\delta <1.
$$ 
Simple algebra yields that
$$
\pr\left(\|\bL\|_{2}^2 < (n_\alpha-r)p \left[\mu_{n_\alpha-r} + \{-2\log(\delta)\}^{1/2} \sigma_{n_\alpha-r}\right]\right) > (1-\delta)^p.
$$ 
Therefore, when the subsample parameter $r$ in Algorithm~1 satisfies $r<\alpha n$ and
\begin{equation}\label{eq:r-general}
\frac{r}{n} \geq \alpha - \frac{\epsilon^{\prime2}\|\X\|_{2}^{2}}{np[\mu_{n_\alpha-r} + \{-2\log(\delta)\}^{1/2} \sigma_{n_\alpha-r}^2 ]},
\end{equation}
it holds that
\begin{equation*}
\pr\left(\|\bL\|_{2}^2 < \epsilon^{\prime2}\|\X\|_{2}^{2} \right) > (1-\delta)^p.
\end{equation*}
Consequently, $\|\X-\X^{*}\|_{2}\leq\epsilon^\prime\|\X\|_{2}$ is achieved with probability at least $(1-\delta)^p$ when $n$ is sufficiently large, for any $0<\delta <1$.

Next, we focus on two specific cases.
\begin{itemize}
    \item Uniform distribution over $(-1,1)$, i.e., $F(x) = 2^{-1}(1+x) \mathbb{I}_{(-1,1)} (x)$, where $\mathbb{I}(\cdot)$ is the indicator function.

    Now we have $G(x) = \sqrt{x} \mathbb{I}_{(0,1)} (x)$ and $g(x) = (2\sqrt{x})^{-1} \mathbb{I}_{(0,1)} (x)$, which leads to $G^{-1}(x) = x^2 \mathbb{I}_{(0,1)} (x)$. It follows that 
    $$
    \mu_{n_\alpha-r} = \left(\frac{n_\alpha-r}{n_\alpha}\right)^2 \quad\text{and}\quad \sigma_{n_\alpha-r}^2 = \frac{4(n_\alpha-r)^3r}{n_\alpha^5}.
    $$
    By plugging above results into~\eqref{eq:r-general}, we conclude that
    \begin{equation*}
    \frac{r}{n} \geq \alpha - \frac{(\alpha \epsilon^{\prime} \|\X\|_{2})^{2/3}}{(2np)^{1/3}},
    \end{equation*}
    for a relatively large $n$. 

    \item Standard normal distribution on $\mathbb{R}$.

    Now the non-zero elements in $\{x_{1 j}^2, \ldots, x_{n j}^2\}$ follow the chi-squared distribution with $1$ degree of freedom. Unfortunately, $\mu_{n_\alpha-r}$ and $\sigma_{n_\alpha-r}^2$ don't have closed-form expressions. According to the assumption $r > n_\alpha (1-\phi)$, it holds that 
    $$
    \mu_{n_\alpha-r} < G^{-1}(\phi) \quad\text{and}\quad \sigma_{n_\alpha-r}^2 < \frac{(n_\alpha-r)r}{n_\alpha^3 g^2\{G^{-1}(\phi)\}}.
    $$
    Hence, when $n$ is sufficiently large and
    \begin{equation*}
    \frac{r}{n} \geq \alpha - \frac{(\epsilon^{\prime} \|\X\|_{2})^2}{2 G^{-1}(\phi) np},
    \end{equation*}
    the inequality~\eqref{eq:r-general} holds.
    
\end{itemize}

\end{proof}

\subsection{Proof of Theorem~\ref{th2:mom}}
\begin{proof}
Since the dimension of the parameter $\bbeta$ is finite and the MOM procedure is adopted coordinate-wisely, without loss of generality, we only prove the univariate case, and we assume the first block contains no outliers.

By the definition of MOM, one can see that
\[\left\{\|\tB_{\rm MOM}-\bbeta\|>\epsilon \right\} \subseteq \left\{\sum_{l=1}^k \mathbb{I}(\|\tB^{(l)}-\bbeta\|>\epsilon)\ge k/2 \right\},\]
where $\mathbb{I}(\cdot)$ is the indicator function and $\tB^{(l)}$ is the core-elements estimator on the $l$th block.
To ease the conversation, denote $Z^{(l)} = \mathbb{I}(\|\tB^{(l)}-\bbeta\|>\epsilon)$ and $\cB_\cO^c$ be the complementary set of $\cB_\cO$, i.e., $\cB_\cO^c = \{1,\ldots,k\} \setminus \cB_\cO$.
Then, we have
\begin{align}
 \nonumber &\pr(\|\tB_{\rm MOM}-\bbeta\|>\epsilon) \\
 \nonumber \le& \pr \left\{\sum_{l=1}^k I(\|\tB^{(l)}-\bbeta\|>\epsilon )\ge k/2 \right\}\\
 \nonumber   =&\pr \left(\sum_{l\in\cB_\cO^c} Z^{(l)} + \sum_{l\in\cB_\cO} Z^{(l)} \ge k/2 \right)\\
 \nonumber   \le& \pr \left(\sum_{l\in\cB_\cO^c} Z^{(l)} +|\cB_\cO|\ge k/2 \right)\\
    =&\pr \left[\frac{1}{k-|\cB_\cO|}\sum_{l\in\cB_\cO^c} \{Z^{(l)} - E(Z^{(l)})\} \ge \frac{k-2|\cB_\cO|}{2(k-|\cB_\cO|)}-E(Z^{(1)})\right]  \label{eq:13}\\
    \le& \exp\left[-2(k-|\cB_\cO|)\left\{\frac{1}{2}-\frac{|\cB_\cO|}{2(k-|\cB_\cO|)}-E(Z^{(1)})\right\}^2\right],\label{eq:14}
\end{align}
where \eqref{eq:13} comes from the fact that the partition is random, so that $\tB^{(l)}$ are i.i.d. for $l\in\cB_\cO^c$, which indicates $E (Z^{(l)}) = E (Z^{(1)})$ for all $l\in\cB_\cO^c$; the inequality \eqref{eq:14} comes from the Hoeffding's inequality.

From Markov's inequality, one can see that
$$E (Z^{(1)}) \le E(\|\tB^{(1)}-\bbeta\|^2)/\epsilon^2\to 0$$
according to Lemma~\ref{lem3} under (H1)--(H3).
Note that ${|\cB_\cO|}/{(2k-2|\cB_\cO|)}\to 0$ since $|\cB_\cO|$ is finite and $k\to \infty$.
Thus the desired result follows.
\end{proof}

\section{Additional numerical results}\label{sec:numerical}
\subsection{Misspecified linear model}\label{sec:mis}
In this section, we show the proposed core-elements estimator is robust to model misspecification. Suppose the true underlying model has the form
\begin{equation}\label{MLM}
y_{i} = \bx_{i}^{\top} \bbeta + u_{i}, \quad i=1, \ldots, n.
\end{equation}
Here, $u_{i}$'s are independently distributed random errors following the non-centered normal distribution $N(h(\bx_{i}), \sigma^{2})$, where $h(\cdot)$ is an unknown multivariate function.
Without prior information on the true model~\eqref{MLM}, the classical linear model~(1) in the manuscript is a misspecified linear model of~\eqref{MLM}.

We simulate the data from the model~\eqref{MLM} with $n = 10^4$, $p=20$ and $r\in\{2p, 4p, 6p, 8p, 10p\}$. The predictor matrix $\X$ is generated from (D1) with different sparsity patterns (R1)--(R5). Other settings are the same as those in subsection~5.1 of the manuscript.
To show the robustness of the proposed method to various misspecification terms, following the work of \cite{meng2021lowcon}, we consider different kinds of $h(\cdot)$:
\begin{enumerate}[(M1)]
    \item $h(\bx_i) = c_1 \cdot x_{i3}x_{i8}$;
    \item $h(\bx_i) = c_2 \cdot x_{i3} \operatorname{sin}(x_{i8})$;
    \item $h(\bx_i) = c_3 \cdot x_{i3}^2$,
\end{enumerate}
where the constants $c_1, c_2$ and $c_3$ are selected so that $\max_{\bx \in\{\bx_i\}_{i=1}^n}|h(\bx)|=10$, that is, the misspecification term does not dominate the response.

\begin{figure}[!t]
    \centering
    \subfigure[Estimation performance.]{\includegraphics[width=5.5in]{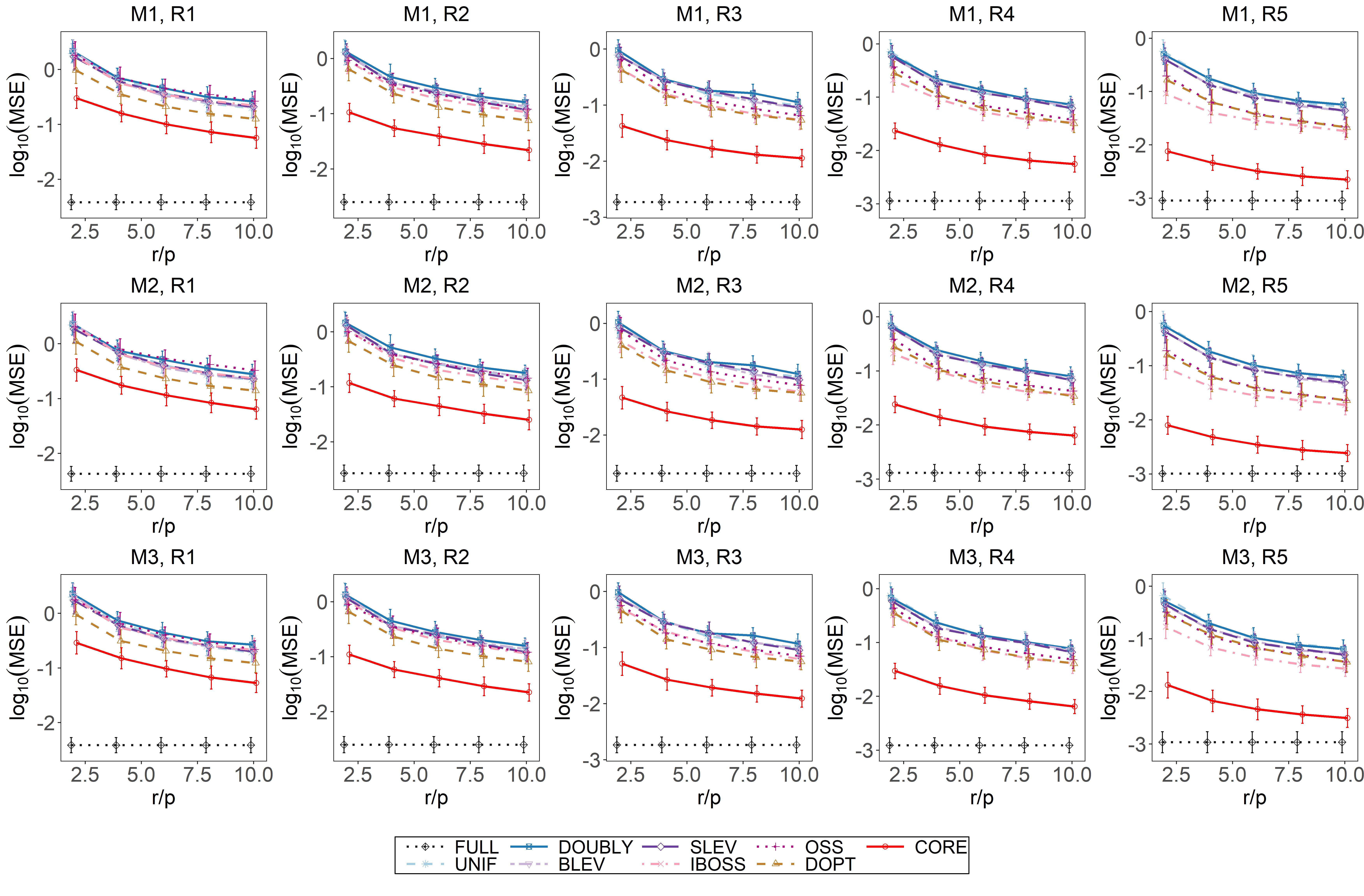}
    \label{fig:simu-mis-mse}}
    \subfigure[Prediction performance.]{\includegraphics[width=5.5in]{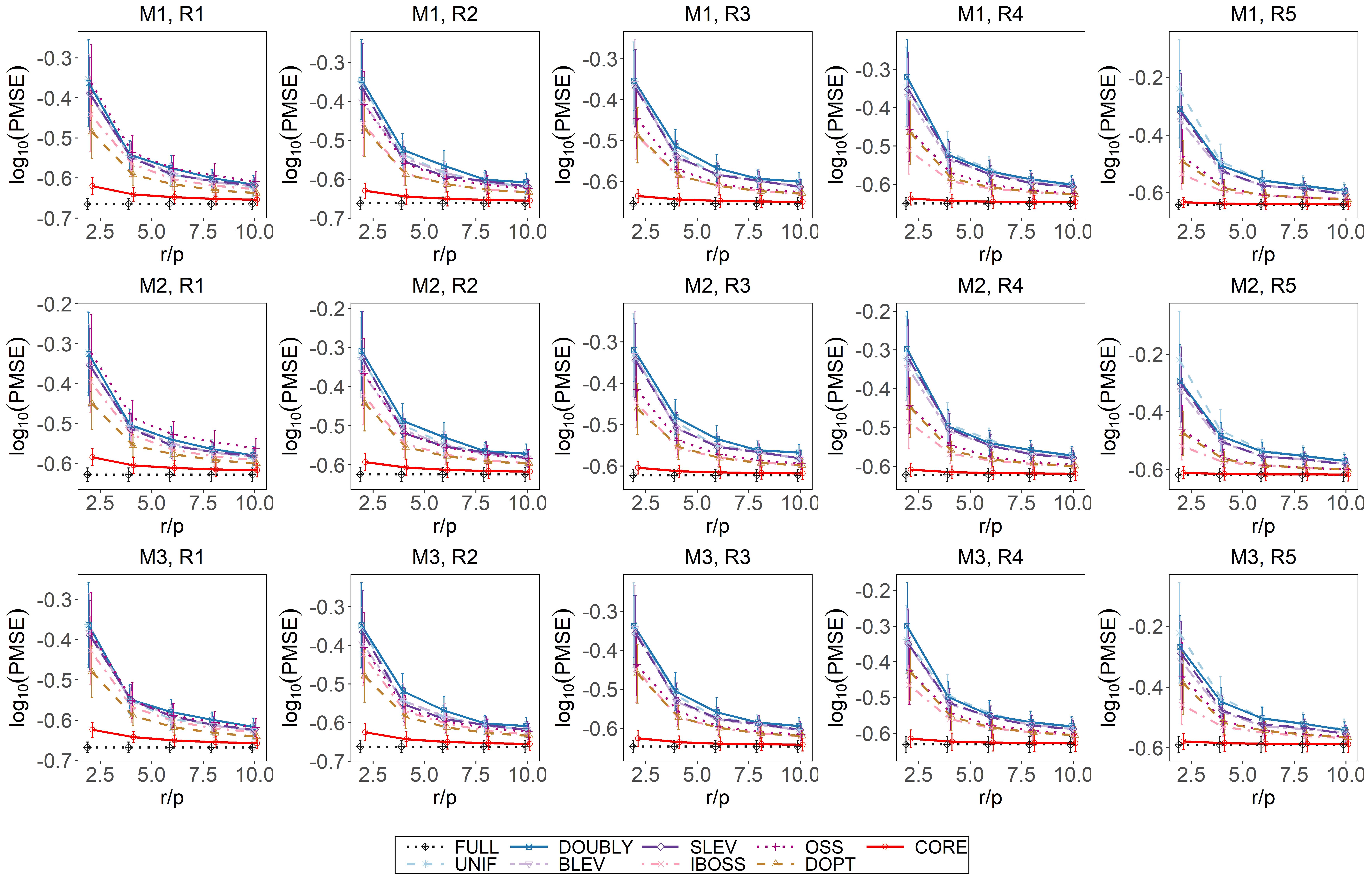}
    \label{fig:simu-mis-pmse}}
    \caption{Comparison of different estimators w.r.t. MSE and PMSE. Each row represents a particular misspecification term, i.e., (M1)--(M3), and each column represents a different sparsity ratio, i.e., (R1)--(R5). Vertical bars are the standard errors.}
    \label{fig:simu-mis}
\end{figure}

Figures~\ref{fig:simu-mis-mse} and~\ref{fig:simu-mis-pmse} compare the performance of different subsample estimators under model misspecification in estimation and prediction, respectively. We observe the proposed core-elements approach yields the most accurate estimation in all cases, indicating the performance of core-elements is robust to model misspecification.

\subsection{Alternative choices of model coefficients}\label{sec:beta}

In subsampling literature, it is common practice to fix the true coefficient $\bbeta$ as a constant vector, as seen in \cite{wang2019information, wang2021orthogonal, yu2023review, chasiotis2024subdata}, and as employed in the main body of our manuscript. However, to comprehensively evaluate the adaptability of our proposed method in different scenarios, we also test it with alternative coefficients, including $\beta_i = (-1)^{i+1}, i=1,\ldots, p$ \citep{hou2023generalized} and $\beta = (1_{10}, 0.1 \times 1_{p-20}, 1_{10})^\top$ \citep{ma2015statistical}. The remaining simulation settings are consistent with those in subsection~5.1.

\begin{figure}[!t]
    \centering
    \subfigure[Estimation performance.]{\includegraphics[width=5.5in]{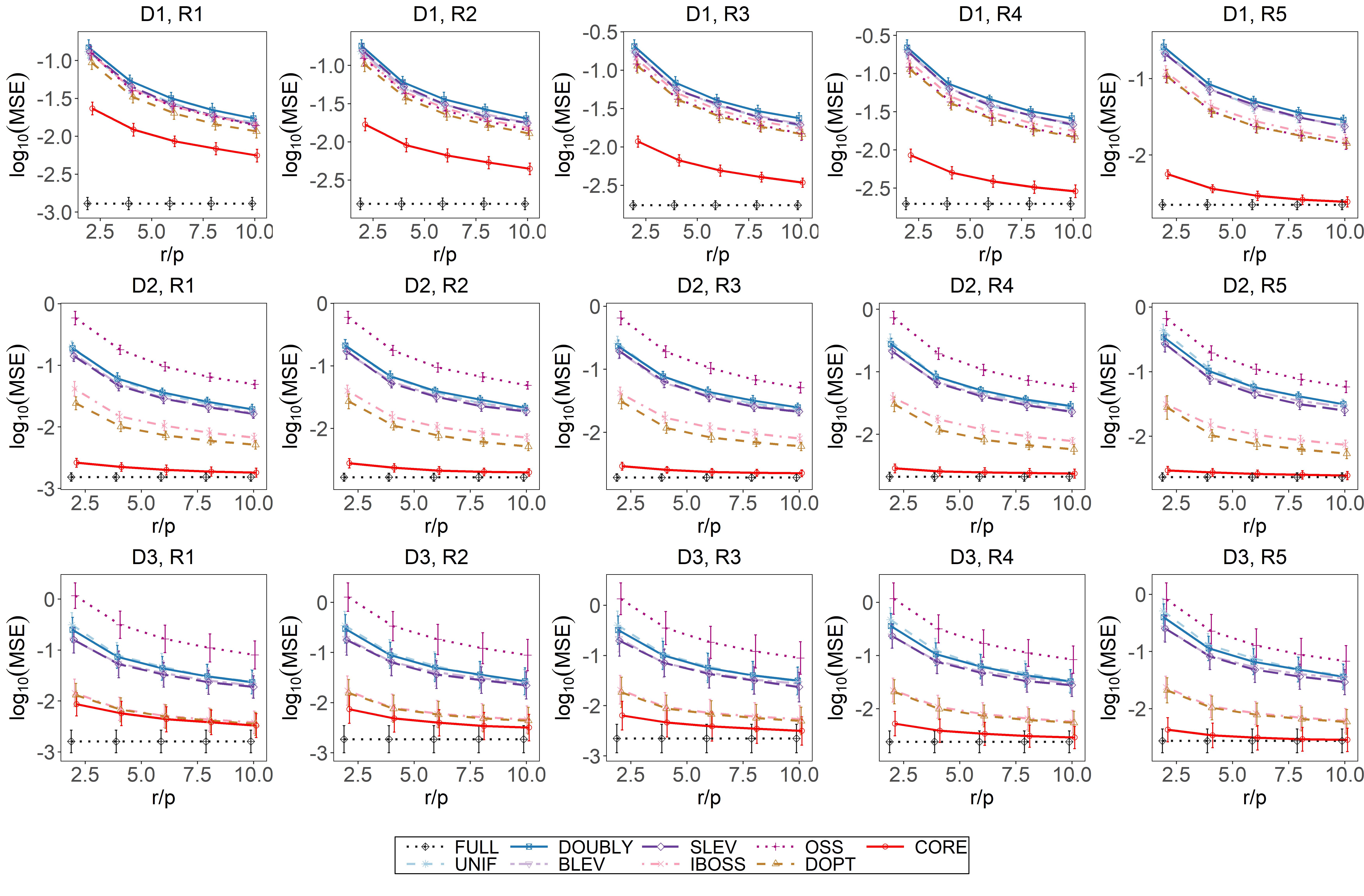}
    \label{fig:simu-beta1-mse}}
    \subfigure[Prediction performance.]{\includegraphics[width=5.5in]{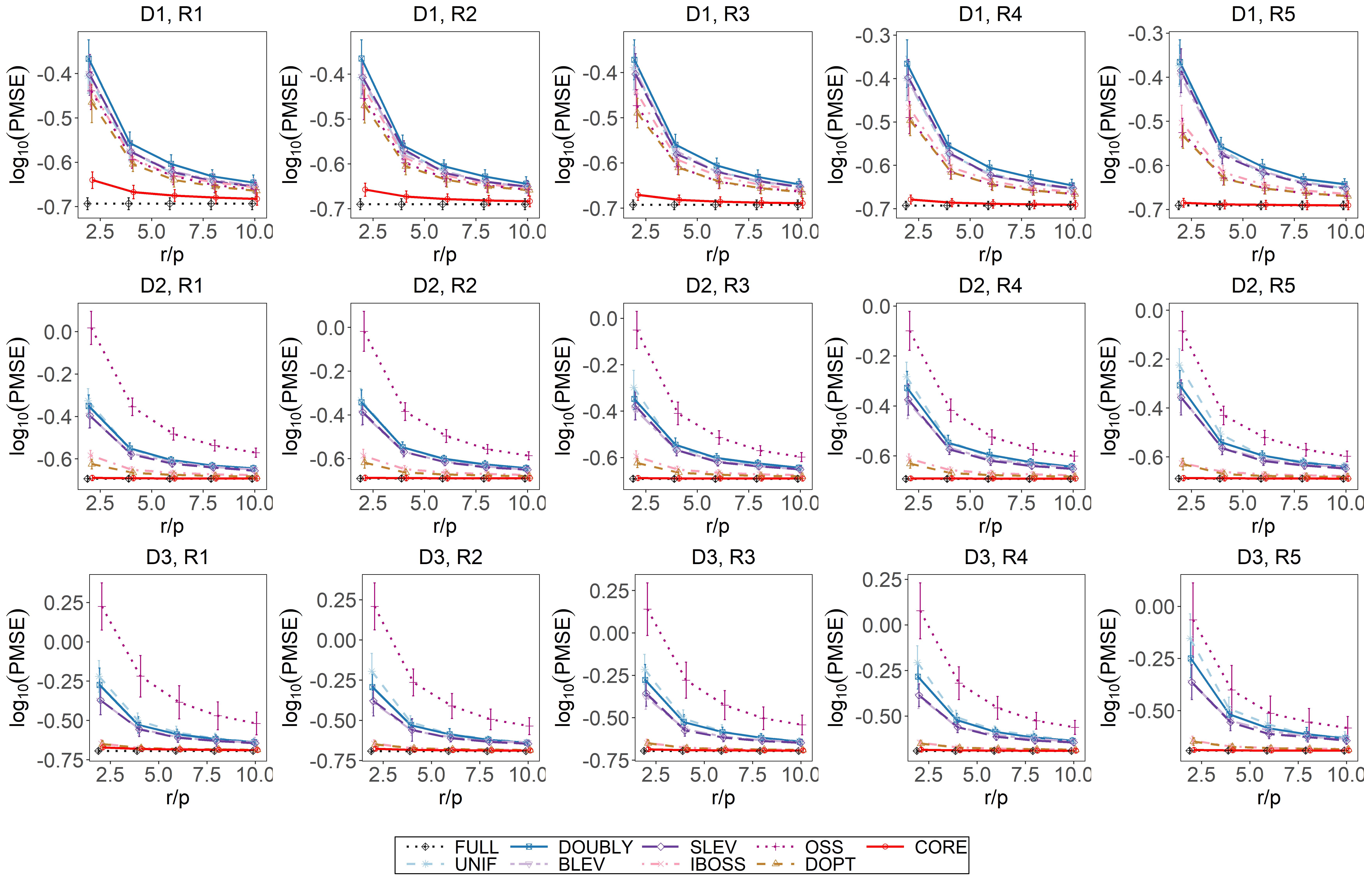}
    \label{fig:simu-beta1-pmse}}
    \caption{Comparison of different estimators w.r.t. MSE and PMSE for $\beta_i = (-1)^{i+1}, i=1,\ldots, p$. Each row represents a particular data distribution, i.e., (D1)--(D3), and each column represents a different sparsity ratio, i.e., (R1)--(R5). Vertical bars are the standard errors.}
    \label{fig:simu-beta1}
\end{figure}

\begin{figure}[!t]
    \centering
    \subfigure[Estimation performance.]{\includegraphics[width=5.5in]{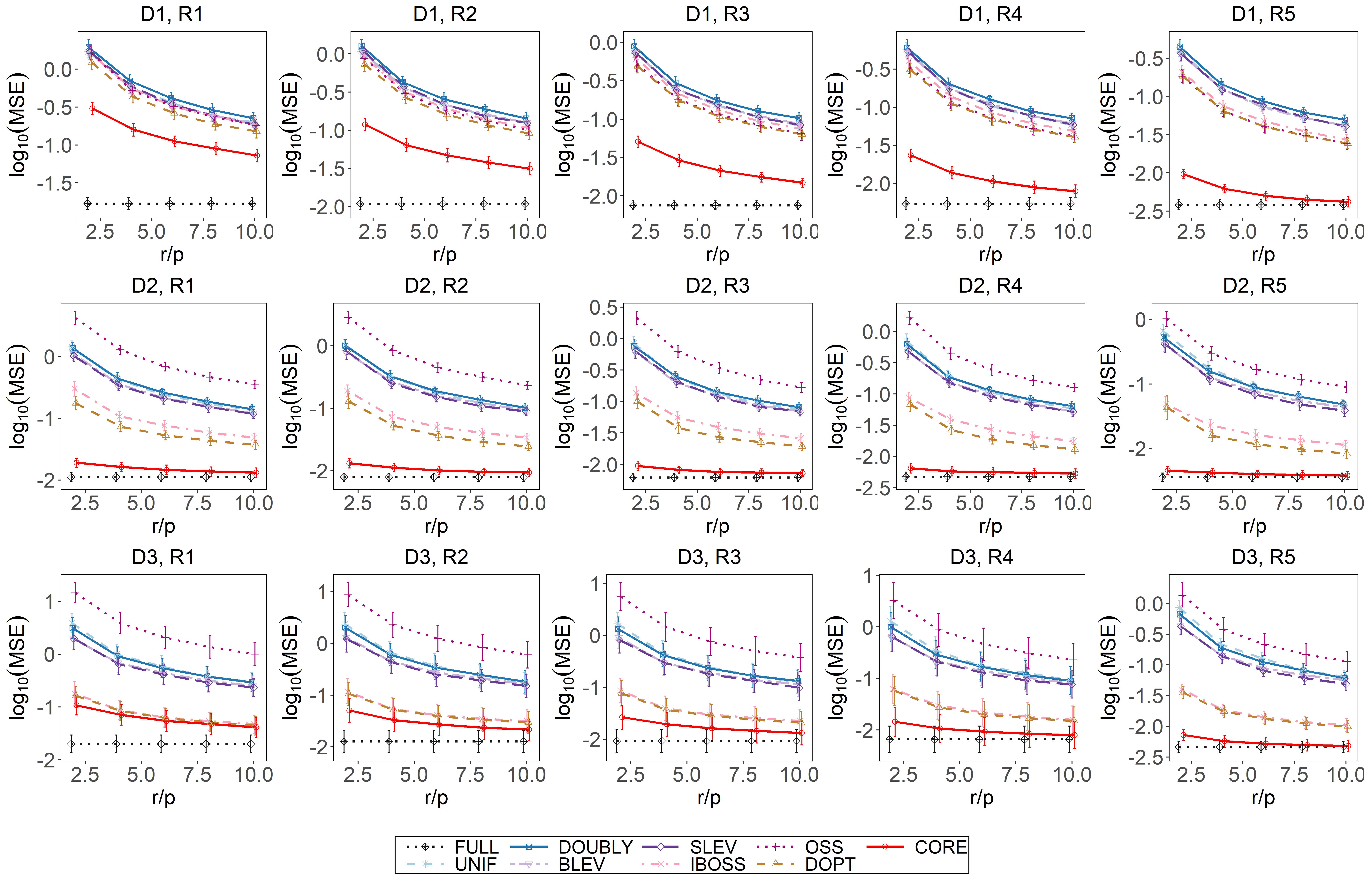}
    \label{fig:simu-beta2-mse}}
    \subfigure[Prediction performance.]{\includegraphics[width=5.5in]{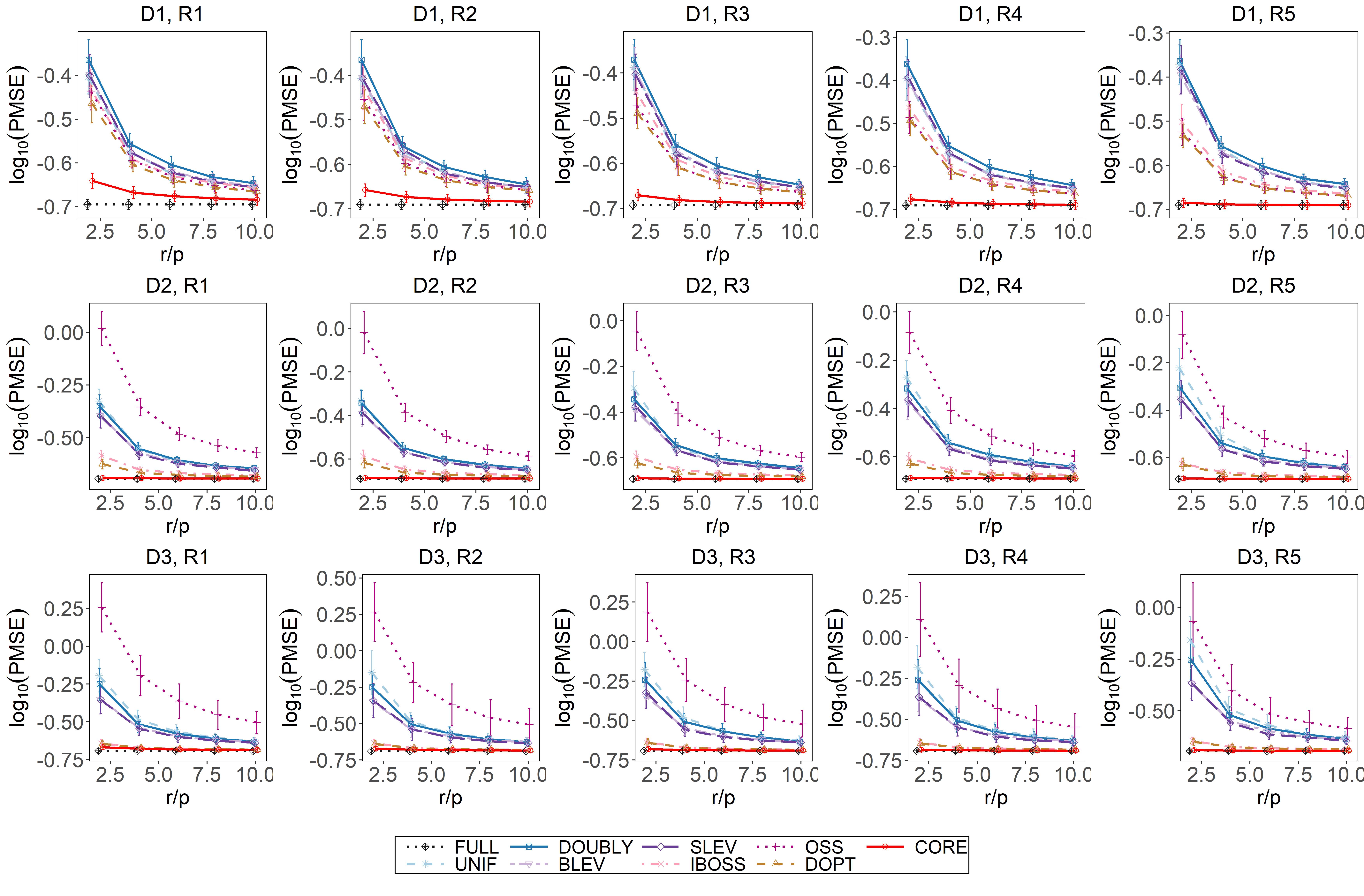}
    \label{fig:simu-beta2-pmse}}
    \caption{Comparison of different estimators w.r.t. MSE and PMSE for $\bbeta = (1_{10}, 0.1 \times 1_{p-20}, 1_{10})^\top$. Each row represents a particular data distribution, i.e., (D1)--(D3), and each column represents a different sparsity ratio, i.e., (R1)--(R5). Vertical bars are the standard errors.}
    \label{fig:simu-beta2}
\end{figure}

The comparative results for the core-elements method and its competitors are presented in Figs.~\ref{fig:simu-beta1} and~\ref{fig:simu-beta2}. In both cases, the performance of the full sample estimator and all subsampling methods show patterns highly similar to those observed in Figs.~2 and~3 of the manuscript, wherein our core-elements method not only surpasses other subsampling approaches but also exhibits comparable efficacy to the full sample estimator, even at a relatively small subsample size (e.g., $r=10p$). These findings highlight the generality and superiority of the core-elements approach across varied parametric landscapes.

\bibliographystyle{chicago}
\bibliography{ref} 

\begin{thebibliography}{}

\bibitem[\protect\citeauthoryear{Achlioptas, Karnin, and Liberty}{Achlioptas
  et~al.}{2013}]{achlioptas2013near}
Achlioptas, D., Z.~S. Karnin, and E.~Liberty (2013).
\newblock Near-optimal entrywise sampling for data matrices.
\newblock {\em Advances in Neural Information Processing Systems\/}~{\em 26},
  1565--1573.

\bibitem[\protect\citeauthoryear{Achlioptas and Mcsherry}{Achlioptas and
  Mcsherry}{2007}]{achlioptas2007fast}
Achlioptas, D. and F.~Mcsherry (2007).
\newblock Fast computation of low-rank matrix approximations.
\newblock {\em Journal of the Association for Computing Machinery\/}~{\em
  54\/}(2), 1--19.

\bibitem[\protect\citeauthoryear{Ai, Wang, Yu, and Zhang}{Ai
  et~al.}{2020}]{ai2020optimal}
Ai, M., F.~Wang, J.~Yu, and H.~Zhang (2020).
\newblock Optimal subsampling for large-scale quantile regression.
\newblock {\em Journal of Complexity\/}~{\em 62}, 101512.

\bibitem[\protect\citeauthoryear{Ai, Yu, Zhang, and Wang}{Ai
  et~al.}{2021}]{ai2021optimal}
Ai, M., J.~Yu, H.~Zhang, and H.~Wang (2021).
\newblock Optimal subsampling algorithms for big data regressions.
\newblock {\em Statistica Sinica\/}~{\em 31}, 749--772.

\bibitem[\protect\citeauthoryear{Alaoui and Mahoney}{Alaoui and
  Mahoney}{2015}]{alaoui2015fast}
Alaoui, A. and M.~W. Mahoney (2015).
\newblock Fast randomized kernel ridge regression with statistical guarantees.
\newblock {\em Advances in Neural Information Processing Systems\/}~{\em 28},
  775--783.

\bibitem[\protect\citeauthoryear{Andrews, Kiselev, McCarthy, and
  Hemberg}{Andrews et~al.}{2021}]{andrews2021tutorial}
Andrews, T.~S., V.~Y. Kiselev, D.~McCarthy, and M.~Hemberg (2021).
\newblock Tutorial: guidelines for the computational analysis of single-cell
  {RNA} sequencing data.
\newblock {\em Nature Protocols\/}~{\em 16\/}(1), 1--9.

\bibitem[\protect\citeauthoryear{Arora, Hazan, and Kale}{Arora
  et~al.}{2005}]{arora2005fast}
Arora, S., E.~Hazan, and S.~Kale (2005).
\newblock Fast algorithms for approximate semidefinite programming using the
  multiplicative weights update method.
\newblock In {\em 46th Annual IEEE Symposium on Foundations of Computer
  Science}, pp.\  339--348. IEEE.

\bibitem[\protect\citeauthoryear{Arora, Hazan, and Kale}{Arora
  et~al.}{2006}]{arora2006fast}
Arora, S., E.~Hazan, and S.~Kale (2006).
\newblock A fast random sampling algorithm for sparsifying matrices.
\newblock In {\em Approximation, Randomization, and Combinatorial Optimization.
  Algorithms and Techniques}, pp.\  272--279. Springer.

\bibitem[\protect\citeauthoryear{Azizi, Carr, Plitas, Cornish, Konopacki,
  Prabhakaran, Nainys, Wu, Kiseliovas, and Setty}{Azizi
  et~al.}{2018}]{azizi2018single}
Azizi, E., A.~J. Carr, G.~Plitas, A.~E. Cornish, C.~Konopacki, S.~Prabhakaran,
  J.~Nainys, K.~Wu, V.~Kiseliovas, and M.~Setty (2018).
\newblock Single-cell map of diverse immune phenotypes in the breast tumor
  microenvironment.
\newblock {\em Cell\/}~{\em 174\/}(5), 1293--1308.

\bibitem[\protect\citeauthoryear{Boutsidis, Drineas, and
  Magdon-Ismail}{Boutsidis et~al.}{2013}]{boutsidis2013near}
Boutsidis, C., P.~Drineas, and M.~Magdon-Ismail (2013).
\newblock Near-optimal coresets for least-squares regression.
\newblock {\em IEEE Transactions on Information Theory\/}~{\em 59\/}(10),
  6880--6892.

\bibitem[\protect\citeauthoryear{Braverman, Krauthgamer, Krishnan, and
  Sapir}{Braverman et~al.}{2021}]{braverman21near}
Braverman, V., R.~Krauthgamer, A.~R. Krishnan, and S.~Sapir (2021).
\newblock Near-optimal entrywise sampling of numerically sparse matrices.
\newblock In {\em Proceedings of Thirty Fourth Conference on Learning Theory},
  Volume 134, pp.\  759--773. PMLR.

\bibitem[\protect\citeauthoryear{Cai, Cand{\`e}s, and Shen}{Cai
  et~al.}{2010}]{cai2010singular}
Cai, J.-F., E.~J. Cand{\`e}s, and Z.~Shen (2010).
\newblock A singular value thresholding algorithm for matrix completion.
\newblock {\em SIAM Journal on Optimization\/}~{\em 20\/}(4), 1956--1982.

\bibitem[\protect\citeauthoryear{Cand{\`e}s and Recht}{Cand{\`e}s and
  Recht}{2009}]{candes2009exact}
Cand{\`e}s, E.~J. and B.~Recht (2009).
\newblock Exact matrix completion via convex optimization.
\newblock {\em Foundations of Computational Mathematics\/}~{\em 9\/}(6),
  717--772.

\bibitem[\protect\citeauthoryear{Cand{\`e}s and Tao}{Cand{\`e}s and
  Tao}{2010}]{candes2010power}
Cand{\`e}s, E.~J. and T.~Tao (2010).
\newblock The power of convex relaxation: Near-optimal matrix completion.
\newblock {\em IEEE Transactions on Information Theory\/}~{\em 56\/}(5),
  2053--2080.

\bibitem[\protect\citeauthoryear{Carmon, Jin, Sidford, and Tian}{Carmon
  et~al.}{2020}]{carmon2020coordinate}
Carmon, Y., Y.~Jin, A.~Sidford, and K.~Tian (2020).
\newblock Coordinate methods for matrix games.
\newblock In {\em 2020 IEEE 61st Annual Symposium on Foundations of Computer
  Science}, pp.\  283--293. IEEE.

\bibitem[\protect\citeauthoryear{Carr, Lewin-Koh, Maechler, and Sarkar}{Carr
  et~al.}{2023}]{carr2023hexbin}
Carr, D., N.~Lewin-Koh, M.~Maechler, and D.~Sarkar (2023).
\newblock hexbin: Hexagonal binning routines.
\newblock {\em R package version 1.28.3\/}.

\bibitem[\protect\citeauthoryear{Chang}{Chang}{2023}]{chang2023predictive}
Chang, M.-C. (2023).
\newblock Predictive subdata selection for computer models.
\newblock {\em Journal of Computational and Graphical Statistics\/}~{\em
  32\/}(2), 613--630.

\bibitem[\protect\citeauthoryear{Chasiotis and Karlis}{Chasiotis and
  Karlis}{2024}]{chasiotis2024subdata}
Chasiotis, V. and D.~Karlis (2024).
\newblock Subdata selection for big data regression: an improved approach.
\newblock {\em Journal of Data Science, Statistics, and Visualisation\/}~{\em
  4\/}(3), 1--28.

\bibitem[\protect\citeauthoryear{Chen, Bhojanapalli, Sanghavi, and Ward}{Chen
  et~al.}{2014}]{chen2014coherent}
Chen, Y., S.~Bhojanapalli, S.~Sanghavi, and R.~Ward (2014).
\newblock Coherent matrix completion.
\newblock In {\em International Conference on Machine Learning}, pp.\
  674--682. PMLR.

\bibitem[\protect\citeauthoryear{Dai, Song, and Wang}{Dai
  et~al.}{2023}]{dai2023subsampling}
Dai, W., Y.~Song, and D.~Wang (2023).
\newblock A subsampling method for regression problems based on minimum energy
  criterion.
\newblock {\em Technometrics\/}~{\em 65\/}(2), 192--205.

\bibitem[\protect\citeauthoryear{Dasgupta, Drineas, Harb, Kumar, and
  Mahoney}{Dasgupta et~al.}{2009}]{dasgupta2009sampling}
Dasgupta, A., P.~Drineas, B.~Harb, R.~Kumar, and M.~W. Mahoney (2009).
\newblock Sampling algorithms and coresets for $l_p$ regression.
\newblock {\em SIAM Journal on Computing\/}~{\em 38\/}(5), 2060--2078.

\bibitem[\protect\citeauthoryear{Davis and Hu}{Davis and
  Hu}{2011}]{davis2011university}
Davis, T.~A. and Y.~Hu (2011).
\newblock The {U}niversity of {F}lorida sparse matrix collection.
\newblock {\em ACM Transactions on Mathematical Software\/}~{\em 38\/}(1),
  1--25.

\bibitem[\protect\citeauthoryear{Derezi{\'n}ski, Warmuth, and
  Hsu}{Derezi{\'n}ski et~al.}{2018}]{derezinski2018leveraged}
Derezi{\'n}ski, M., M.~K. Warmuth, and D.~J. Hsu (2018).
\newblock Leveraged volume sampling for linear regression.
\newblock {\em Advances in Neural Information Processing Systems\/}~{\em 31},
  2510--2519.

\bibitem[\protect\citeauthoryear{Donoho and Gasko}{Donoho and
  Gasko}{1992}]{donoho1992breakdown}
Donoho, D.~L. and M.~Gasko (1992).
\newblock Breakdown properties of location estimates based on halfspace depth
  and projected outlyingness.
\newblock {\em The Annals of Statistics\/}~{\em 20\/}(4), 1803--1827.

\bibitem[\protect\citeauthoryear{Donoho and Huber}{Donoho and
  Huber}{1983}]{donoho1983notion}
Donoho, D.~L. and P.~J. Huber (1983).
\newblock The notion of breakdown point.
\newblock {\em A Festschrift for Erich L. Lehmann\/}.

\bibitem[\protect\citeauthoryear{Drineas, Kannan, and Mahoney}{Drineas
  et~al.}{2006}]{drineas2006fast}
Drineas, P., R.~Kannan, and M.~W. Mahoney (2006).
\newblock Fast {M}onte {C}arlo algorithms for matrices {I}: Approximating
  matrix multiplication.
\newblock {\em SIAM Journal on Computing\/}~{\em 36\/}(1), 132--157.

\bibitem[\protect\citeauthoryear{Drineas, Magdon-Ismail, Mahoney, and
  Woodruff}{Drineas et~al.}{2012}]{drineas2012fast}
Drineas, P., M.~Magdon-Ismail, M.~W. Mahoney, and D.~P. Woodruff (2012).
\newblock Fast approximation of matrix coherence and statistical leverage.
\newblock {\em The Journal of Machine Learning Research\/}~{\em 13\/}(1),
  3475--3506.

\bibitem[\protect\citeauthoryear{Drineas and Zouzias}{Drineas and
  Zouzias}{2011}]{drineas2011note}
Drineas, P. and A.~Zouzias (2011).
\newblock A note on element-wise matrix sparsification via a matrix-valued
  {B}ernstein inequality.
\newblock {\em Information Processing Letters\/}~{\em 111\/}(8), 385--389.

\bibitem[\protect\citeauthoryear{d’Aspremont}{d’Aspremont}{2011}]{d2011subsampling}
d’Aspremont, A. (2011).
\newblock Subsampling algorithms for semidefinite programming.
\newblock {\em Stochastic Systems\/}~{\em 1\/}(2), 274--305.

\bibitem[\protect\citeauthoryear{Edgar, Domrachev, and Lash}{Edgar
  et~al.}{2002}]{edgar2002gene}
Edgar, R., M.~Domrachev, and A.~E. Lash (2002).
\newblock Gene {E}xpression {O}mnibus: {NCBI} gene expression and hybridization
  array data repository.
\newblock {\em Nucleic Acids Research\/}~{\em 30\/}(1), 207--210.

\bibitem[\protect\citeauthoryear{El~Karoui and d’Aspremont}{El~Karoui and
  d’Aspremont}{2010}]{el2010second}
El~Karoui, N. and A.~d’Aspremont (2010).
\newblock Second order accurate distributed eigenvector computation for
  extremely large matrices.
\newblock {\em Electronic Journal of Statistics\/}~{\em 4}, 1345--1385.

\bibitem[\protect\citeauthoryear{Feldman, Schmidt, and Sohler}{Feldman
  et~al.}{2020}]{feldman2020turning}
Feldman, D., M.~Schmidt, and C.~Sohler (2020).
\newblock Turning big data into tiny data: Constant-size coresets for
  $k$-means, {PCA}, and projective clustering.
\newblock {\em SIAM Journal on Computing\/}~{\em 49\/}(3), 601--657.

\bibitem[\protect\citeauthoryear{Garber and Hazan}{Garber and
  Hazan}{2016}]{garber2016sublinear}
Garber, D. and E.~Hazan (2016).
\newblock Sublinear time algorithms for approximate semidefinite programming.
\newblock {\em Mathematical Programming\/}~{\em 158\/}(1), 329--361.

\bibitem[\protect\citeauthoryear{Gupta and Sidford}{Gupta and
  Sidford}{2018}]{gupta2018exploiting}
Gupta, N. and A.~Sidford (2018).
\newblock Exploiting numerical sparsity for efficient learning: faster
  eigenvector computation and regression.
\newblock {\em Advances in Neural Information Processing Systems\/}~{\em 31},
  5274--5283.

\bibitem[\protect\citeauthoryear{Hager}{Hager}{1989}]{hager1989updating}
Hager, W.~W. (1989).
\newblock Updating the inverse of a matrix.
\newblock {\em SIAM Review\/}~{\em 31\/}(2), 221--239.

\bibitem[\protect\citeauthoryear{Hampel}{Hampel}{1968}]{hampel1968contributions}
Hampel, F.~R. (1968).
\newblock {\em Contributions to the Theory of Robust Estimation}.
\newblock University of California, Berkeley.

\bibitem[\protect\citeauthoryear{Hastie, Mazumder, Lee, and Zadeh}{Hastie
  et~al.}{2015}]{hastie2015matrix}
Hastie, T., R.~Mazumder, J.~D. Lee, and R.~Zadeh (2015).
\newblock Matrix completion and low-rank {SVD} via fast alternating least
  squares.
\newblock {\em Journal of Machine Learning Research\/}~{\em 16\/}(1),
  3367--3402.

\bibitem[\protect\citeauthoryear{Higham}{Higham}{2008}]{higham2008functions}
Higham, N.~J. (2008).
\newblock {\em Functions of Matrices: Theory and Computation}.
\newblock SIAM.

\bibitem[\protect\citeauthoryear{Hou-Liu and Browne}{Hou-Liu and
  Browne}{2023}]{hou2023generalized}
Hou-Liu, J. and R.~P. Browne (2023).
\newblock Generalized linear models for massive data via doubly-sketching.
\newblock {\em Statistics and Computing\/}~{\em 33\/}(105), 1--19.

\bibitem[\protect\citeauthoryear{Hsu and Sabato}{Hsu and
  Sabato}{2014}]{hsu2014heavy}
Hsu, D. and S.~Sabato (2014).
\newblock Heavy-tailed regression with a generalized median-of-means.
\newblock In {\em International Conference on Machine Learning}, pp.\  37--45.
  PMLR.

\bibitem[\protect\citeauthoryear{Hu, Li, Liu, and Meng}{Hu
  et~al.}{2024}]{hu2024sampling}
Hu, Y., M.~Li, X.~Liu, and C.~Meng (2024).
\newblock Sampling-based methods for multi-block optimization problems over
  transport polytopes.
\newblock {\em Mathematics of Computation\/}, In press.

\bibitem[\protect\citeauthoryear{Huang, Ma, and Zhang}{Huang
  et~al.}{2008}]{huang2008adaptive}
Huang, J., S.~Ma, and C.~H. Zhang (2008).
\newblock Adaptive {L}asso for sparse high-dimensional regression models.
\newblock {\em Statistica Sinica\/}~{\em 18\/}(4), 1603--1618.

\bibitem[\protect\citeauthoryear{Huang and Lederer}{Huang and
  Lederer}{2023}]{huang2023deepmom}
Huang, S.-T. and J.~Lederer (2023).
\newblock Deepmom: Robust deep learning with median-of-means.
\newblock {\em Journal of Computational and Graphical Statistics\/}~{\em
  32\/}(1), 181--195.

\bibitem[\protect\citeauthoryear{Joseph and Mak}{Joseph and
  Mak}{2021}]{joseph2021supervised}
Joseph, V.~R. and S.~Mak (2021).
\newblock Supervised compression of big data.
\newblock {\em Statistical Analysis and Data Mining: The ASA Data Science
  Journal\/}~{\em 14\/}(3), 217--229.

\bibitem[\protect\citeauthoryear{Joseph and Vakayil}{Joseph and
  Vakayil}{2022}]{joseph2022split}
Joseph, V.~R. and A.~Vakayil (2022).
\newblock {SP}lit: An optimal method for data splitting.
\newblock {\em Technometrics\/}~{\em 64\/}(2), 166--176.

\bibitem[\protect\citeauthoryear{Kairouz, McMahan, Avent, Bellet, Bennis,
  Bhagoji, Bonawitz, Charles, Cormode, Cummings, et~al.}{Kairouz
  et~al.}{2021}]{kairouz2021advances}
Kairouz, P., H.~B. McMahan, B.~Avent, A.~Bellet, M.~Bennis, A.~N. Bhagoji,
  K.~Bonawitz, Z.~Charles, G.~Cormode, R.~Cummings, et~al. (2021).
\newblock Advances and open problems in federated learning.
\newblock {\em Foundations and Trends{\textregistered} in Machine
  Learning\/}~{\em 14\/}(1--2), 1--210.

\bibitem[\protect\citeauthoryear{Knight}{Knight}{2018}]{knight2018subsampling}
Knight, K. (2018).
\newblock Subsampling least squares and elemental estimation.
\newblock In {\em 2018 IEEE Data Science Workshop (DSW)}, pp.\  91--94. IEEE.

\bibitem[\protect\citeauthoryear{Kone{\v{c}}n{\`y}, McMahan, Yu, Richt{\'a}rik,
  Suresh, and Bacon}{Kone{\v{c}}n{\`y} et~al.}{2016}]{konevcny2016federated}
Kone{\v{c}}n{\`y}, J., H.~B. McMahan, F.~X. Yu, P.~Richt{\'a}rik, A.~T. Suresh,
  and D.~Bacon (2016).
\newblock Federated learning: Strategies for improving communication
  efficiency.
\newblock In {\em NIPS Workshop on Private Multi-Party Machine Learning}.

\bibitem[\protect\citeauthoryear{Kundu, Drineas, and Magdon-Ismail}{Kundu
  et~al.}{2017}]{kundu2017recovering}
Kundu, A., P.~Drineas, and M.~Magdon-Ismail (2017).
\newblock Recovering {PCA} and sparse {PCA} via hybrid-$(\ell_1, \ell_2)$
  sparse sampling of data elements.
\newblock {\em Journal of Machine Learning Research\/}~{\em 18\/}(1),
  2558--2591.

\bibitem[\protect\citeauthoryear{Lecu{\'e} and Lerasle}{Lecu{\'e} and
  Lerasle}{2019}]{lecue2019learning}
Lecu{\'e}, G. and M.~Lerasle (2019).
\newblock Learning from {MOM}’s principles: Le {C}am’s approach.
\newblock {\em Stochastic Processes and Their Applications\/}~{\em 129\/}(11),
  4385--4410.

\bibitem[\protect\citeauthoryear{Lecu{\'e} and Lerasle}{Lecu{\'e} and
  Lerasle}{2020}]{lecue2020robust}
Lecu{\'e}, G. and M.~Lerasle (2020).
\newblock Robust machine learning by median-of-means: theory and practice.
\newblock {\em The Annals of Statistics\/}~{\em 48\/}(2), 906--931.

\bibitem[\protect\citeauthoryear{Li, Xie, Wang, Guo, Ye, Ma, and Song}{Li
  et~al.}{2019}]{li2019online}
Li, F., R.~Xie, Z.~Wang, L.~Guo, J.~Ye, P.~Ma, and W.~Song (2019).
\newblock Online distributed {I}o{T} security monitoring with multidimensional
  streaming big data.
\newblock {\em IEEE Internet of Things Journal\/}~{\em 7\/}(5), 4387--4394.

\bibitem[\protect\citeauthoryear{Li, Yu, Li, and Meng}{Li
  et~al.}{023a}]{li2023importance}
Li, M., J.~Yu, T.~Li, and C.~Meng (2023a).
\newblock Importance sparsification for {S}inkhorn algorithm.
\newblock {\em Journal of Machine Learning Research\/}~{\em 24}, 1--44.

\bibitem[\protect\citeauthoryear{Li, Yu, Xu, and Meng}{Li
  et~al.}{023b}]{li2023efficient}
Li, M., J.~Yu, H.~Xu, and C.~Meng (2023b).
\newblock Efficient approximation of {G}romov-{W}asserstein distance using
  importance sparsification.
\newblock {\em Journal of Computational and Graphical Statistics\/}~{\em
  32\/}(4), 1512--1523.

\bibitem[\protect\citeauthoryear{Li, Zhang, and Meng}{Li
  et~al.}{2024}]{li2024nonparametric}
Li, M., J.~Zhang, and C.~Meng (2024).
\newblock Nonparametric additive models for billion observations.
\newblock {\em Journal of Computational and Graphical Statistics\/}, In press.

\bibitem[\protect\citeauthoryear{Li and Meng}{Li and Meng}{2021}]{li2021modern}
Li, T. and C.~Meng (2021).
\newblock Modern subsampling methods for large-scale least squares regression.
\newblock {\em International Journal of Cyber-Physical Systems\/}~{\em 2\/}(2),
  1--28.

\bibitem[\protect\citeauthoryear{Liu, Zhu, Yuan, Li, and Xue}{Liu
  et~al.}{2021}]{liu2021privacy}
Liu, F., B.~Zhu, S.~Yuan, J.~Li, and K.~Xue (2021).
\newblock Privacy-preserving truth discovery for sparse data in mobile
  crowdsensing systems.
\newblock In {\em 2021 IEEE Global Communications Conference}, pp.\  1--6.
  IEEE.

\bibitem[\protect\citeauthoryear{Lugosi and Mendelson}{Lugosi and
  Mendelson}{2019}]{lugosi2019regularization}
Lugosi, G. and S.~Mendelson (2019).
\newblock Regularization, sparse recovery, and median-of-means tournaments.
\newblock {\em Bernoulli\/}~{\em 25\/}(3), 2075--2106.

\bibitem[\protect\citeauthoryear{Ma, Huang, and Zhang}{Ma
  et~al.}{2015}]{ma2015efficient}
Ma, P., J.~Z. Huang, and N.~Zhang (2015).
\newblock Efficient computation of smoothing splines via adaptive basis
  sampling.
\newblock {\em Biometrika\/}~{\em 102\/}(3), 631--645.

\bibitem[\protect\citeauthoryear{Ma, Mahoney, and Yu}{Ma
  et~al.}{2015}]{ma2015statistical}
Ma, P., M.~W. Mahoney, and B.~Yu (2015).
\newblock A statistical perspective on algorithmic leveraging.
\newblock {\em Journal of Machine Learning Research\/}~{\em 16\/}(1), 861--911.

\bibitem[\protect\citeauthoryear{Ma and Sun}{Ma and
  Sun}{2015}]{ma2015leveraging}
Ma, P. and X.~Sun (2015).
\newblock Leveraging for big data regression.
\newblock {\em Wiley Interdisciplinary Reviews: Computational
  Statistics\/}~{\em 7\/}(1), 70--76.

\bibitem[\protect\citeauthoryear{Ma, Zhang, Xing, Ma, and Mahoney}{Ma
  et~al.}{2020}]{ma2020asymptotic}
Ma, P., X.~Zhang, X.~Xing, J.~Ma, and M.~Mahoney (2020).
\newblock Asymptotic analysis of sampling estimators for randomized numerical
  linear algebra algorithms.
\newblock In {\em International Conference on Artificial Intelligence and
  Statistics}, pp.\  1026--1035. PMLR.

\bibitem[\protect\citeauthoryear{Maalouf, Eini, Mussay, Feldman, and
  Osadchy}{Maalouf et~al.}{2022}]{maalouf2022unified}
Maalouf, A., G.~Eini, B.~Mussay, D.~Feldman, and M.~Osadchy (2022).
\newblock A unified approach to coreset learning.
\newblock {\em IEEE Transactions on Neural Networks and Learning Systems\/},
  1--13.

\bibitem[\protect\citeauthoryear{Mak and Joseph}{Mak and
  Joseph}{2018}]{mak2018support}
Mak, S. and V.~R. Joseph (2018).
\newblock Support points.
\newblock {\em The Annals of Statistics\/}~{\em 46\/}(6A), 2562--2592.

\bibitem[\protect\citeauthoryear{Mart{\i}nez}{Mart{\i}nez}{2004}]{martinez2004partial}
Mart{\i}nez, C. (2004).
\newblock Partial quicksort.
\newblock In {\em Proc. 6th ACMSIAM Workshop on Algorithm Engineering and
  Experiments and 1st ACM-SIAM Workshop on Analytic Algorithmics and
  Combinatorics}, pp.\  224--228.

\bibitem[\protect\citeauthoryear{Mathieu}{Mathieu}{2021}]{mathieu2021m}
Mathieu, T. (2021).
\newblock {\em M-estimation and Median of Means applied to statistical
  learning}.
\newblock Ph.\ D. thesis, Universit{\'e} Paris-Saclay.

\bibitem[\protect\citeauthoryear{Meng, Wang, Zhang, Mandal, Zhong, and Ma}{Meng
  et~al.}{2017}]{meng2017effective}
Meng, C., Y.~Wang, X.~Zhang, A.~Mandal, W.~Zhong, and P.~Ma (2017).
\newblock Effective statistical methods for big data analytics.
\newblock In {\em Handbook of Research on Applied Cybernetics and Systems
  Science}, pp.\  280--299. IGI Global.

\bibitem[\protect\citeauthoryear{Meng, Xie, Mandal, Zhang, Zhong, and Ma}{Meng
  et~al.}{2021}]{meng2021lowcon}
Meng, C., R.~Xie, A.~Mandal, X.~Zhang, W.~Zhong, and P.~Ma (2021).
\newblock Lowcon: A design-based subsampling approach in a misspecified linear
  model.
\newblock {\em Journal of Computational and Graphical Statistics\/}~{\em
  30\/}(3), 694--708.

\bibitem[\protect\citeauthoryear{Meng, Yu, Chen, Zhong, and Ma}{Meng
  et~al.}{2022}]{meng2022smoothing}
Meng, C., J.~Yu, Y.~Chen, W.~Zhong, and P.~Ma (2022).
\newblock Smoothing splines approximation using {H}ilbert curve basis
  selection.
\newblock {\em Journal of Computational and Graphical Statistics\/}, 1--11.

\bibitem[\protect\citeauthoryear{Meng, Zhang, Zhang, Zhong, and Ma}{Meng
  et~al.}{2020}]{meng2020more}
Meng, C., X.~Zhang, J.~Zhang, W.~Zhong, and P.~Ma (2020).
\newblock More efficient approximation of smoothing splines via space-filling
  basis selection.
\newblock {\em Biometrika\/}~{\em 107}, 723--735.

\bibitem[\protect\citeauthoryear{Munteanu, Schwiegelshohn, Sohler, and
  Woodruff}{Munteanu et~al.}{2018}]{munteanu2018coresets}
Munteanu, A., C.~Schwiegelshohn, C.~Sohler, and D.~P. Woodruff (2018).
\newblock On coresets for logistic regression.
\newblock {\em Advances in Neural Information Processing Systems\/}~{\em 31},
  6562--6571.

\bibitem[\protect\citeauthoryear{Musser}{Musser}{1997}]{musser1997introspective}
Musser, D.~R. (1997).
\newblock Introspective sorting and selection algorithms.
\newblock {\em Software: Practice and Experience\/}~{\em 27\/}(8), 983--993.

\bibitem[\protect\citeauthoryear{Muzellec, Josse, Boyer, and Cuturi}{Muzellec
  et~al.}{2020}]{muzellec2020missing}
Muzellec, B., J.~Josse, C.~Boyer, and M.~Cuturi (2020).
\newblock Missing data imputation using optimal transport.
\newblock In {\em International Conference on Machine Learning}, pp.\
  7130--7140. PMLR.

\bibitem[\protect\citeauthoryear{Nguyen, Van~den Berge, Chiogna, and
  Risso}{Nguyen et~al.}{2023}]{nguyen2023structure}
Nguyen, T. K.~H., K.~Van~den Berge, M.~Chiogna, and D.~Risso (2023).
\newblock Structure learning for zero-inflated counts with an application to
  single-cell {RNA} sequencing data.
\newblock {\em The Annals of Applied Statistics\/}~{\em 17\/}(3), 2555--2573.

\bibitem[\protect\citeauthoryear{Qaiser and Ali}{Qaiser and
  Ali}{2018}]{qaiser2018text}
Qaiser, S. and R.~Ali (2018).
\newblock Text mining: Use of {TF}-{IDF} to examine the relevance of words to
  documents.
\newblock {\em International Journal of Computer Applications\/}~{\em
  181\/}(1), 25--29.

\bibitem[\protect\citeauthoryear{Reuter and Schwabe}{Reuter and
  Schwabe}{2024}]{reuter2023d}
Reuter, T. and R.~Schwabe (2024).
\newblock D-optimal subsampling design for massive data linear regression.
\newblock {\em arXiv preprint arXiv:2307.02236\/}.

\bibitem[\protect\citeauthoryear{Settles}{Settles}{2012}]{settles2012active}
Settles, B. (2012).
\newblock Active learning.
\newblock {\em Synthesis Lectures on Artificial Intelligence and Machine
  Learning\/}~{\em 6\/}(1), 1--114.

\bibitem[\protect\citeauthoryear{Sun, Zhong, and Ma}{Sun
  et~al.}{2021}]{sun2021asymptotic}
Sun, X., W.~Zhong, and P.~Ma (2021).
\newblock An asymptotic and empirical smoothing parameters selection method for
  smoothing spline anova models in large samples.
\newblock {\em Biometrika\/}~{\em 108\/}(1), 149--166.

\bibitem[\protect\citeauthoryear{Vakayil and Joseph}{Vakayil and
  Joseph}{2022}]{vakayil2022data}
Vakayil, A. and V.~R. Joseph (2022).
\newblock Data twinning.
\newblock {\em Statistical Analysis and Data Mining: The ASA Data Science
  Journal\/}~{\em 15\/}(5), 598--610.

\bibitem[\protect\citeauthoryear{Van~Buuren and Groothuis-Oudshoorn}{Van~Buuren
  and Groothuis-Oudshoorn}{2011}]{van2011mice}
Van~Buuren, S. and K.~Groothuis-Oudshoorn (2011).
\newblock {MICE}: Multivariate imputation by chained equations in {R}.
\newblock {\em Journal of Statistical Software\/}~{\em 45}, 1--67.

\bibitem[\protect\citeauthoryear{Walker}{Walker}{1968}]{walker1968note}
Walker, A. (1968).
\newblock A note on the asymptotic distribution of sample quantiles.
\newblock {\em Journal of the Royal Statistical Society Series B: Statistical
  Methodology\/}~{\em 30\/}(3), 570--575.

\bibitem[\protect\citeauthoryear{Wang, Yang, and Stufken}{Wang
  et~al.}{2019}]{wang2019information}
Wang, H., M.~Yang, and J.~Stufken (2019).
\newblock Information-based optimal subdata selection for big data linear
  regression.
\newblock {\em Journal of the American Statistical Association\/}~{\em
  114\/}(525), 393--405.

\bibitem[\protect\citeauthoryear{Wang, Zhu, and Ma}{Wang
  et~al.}{2018}]{wang2018optimal}
Wang, H., R.~Zhu, and P.~Ma (2018).
\newblock Optimal subsampling for large sample logistic regression.
\newblock {\em Journal of the American Statistical Association\/}~{\em
  113\/}(522), 829--844.

\bibitem[\protect\citeauthoryear{Wang, Zou, and Wang}{Wang
  et~al.}{2022}]{wang2022sampling}
Wang, J., J.~Zou, and H.~Wang (2022).
\newblock Sampling with replacement vs {P}oisson sampling: a comparative study
  in optimal subsampling.
\newblock {\em IEEE Transactions on Information Theory\/}~{\em 68\/}(10),
  6605--6630.

\bibitem[\protect\citeauthoryear{Wang, Elmstedt, Wong, and Xu}{Wang
  et~al.}{2021}]{wang2021orthogonal}
Wang, L., J.~Elmstedt, W.~K. Wong, and H.~Xu (2021).
\newblock Orthogonal subsampling for big data linear regression.
\newblock {\em The Annals of Applied Statistics\/}~{\em 15\/}(3), 1273--1290.

\bibitem[\protect\citeauthoryear{Wang, Ouyang, Panpan, and Xu}{Wang
  et~al.}{2023}]{wang2023fast}
Wang, R., Y.~Ouyang, Y.~Panpan, and W.~Xu (2023).
\newblock A fast and accurate estimator for large scale linear model via data
  averaging.
\newblock In {\em Advances in Neural Information Processing Systems},
  Volume~36, pp.\  34917--34927.

\bibitem[\protect\citeauthoryear{Wang, Gittens, and Mahoney}{Wang
  et~al.}{2019}]{wang2019scalable}
Wang, S., A.~Gittens, and M.~W. Mahoney (2019).
\newblock Scalable kernel {K}-means clustering with {N}ystr{\"o}m
  approximation: relative-error bounds.
\newblock {\em Journal of Machine Learning Research\/}~{\em 20\/}(1), 431--479.

\bibitem[\protect\citeauthoryear{Wang and Zhang}{Wang and
  Zhang}{2013}]{wang2013improving}
Wang, S. and Z.~Zhang (2013).
\newblock Improving {CUR} matrix decomposition and the {N}ystr{\"o}m
  approximation via adaptive sampling.
\newblock {\em Journal of Machine Learning Research\/}~{\em 14\/}(1),
  2729--2769.

\bibitem[\protect\citeauthoryear{Xie, Wang, Bai, Ma, and Zhong}{Xie
  et~al.}{2019}]{xie2019online}
Xie, R., Z.~Wang, S.~Bai, P.~Ma, and W.~Zhong (2019).
\newblock Online decentralized leverage score sampling for streaming
  multidimensional time series.
\newblock In {\em The 22nd International Conference on Artificial Intelligence
  and Statistics}, pp.\  2301--2311.

\bibitem[\protect\citeauthoryear{Yu, Ai, and Ye}{Yu
  et~al.}{2023}]{yu2023review}
Yu, J., M.~Ai, and Z.~Ye (2023).
\newblock A review on design inspired subsampling for big data.
\newblock {\em Statistical Papers\/}, 1--44.

\bibitem[\protect\citeauthoryear{Yu, Wang, Ai, and Zhang}{Yu
  et~al.}{2022}]{yu2022optimal}
Yu, J., H.~Wang, M.~Ai, and H.~Zhang (2022).
\newblock Optimal distributed subsampling for maximum quasi-likelihood
  estimators with massive data.
\newblock {\em Journal of the American Statistical Association\/}~{\em
  117\/}(537), 265--276.

\end{thebibliography}

\end{document}